# Authentication and Integrity of Smartphone Videos through Multimedia Container Structure Analysis

Carlos Quinto Huamán[a], Ana Lucila Sandoval Orozco[a], Luis Javier García Villalba[a,*]

[a]*Group of Analysis, Security and Systems (GASS)*
*Department of Software Engineering and Artificial Intelligence (DISIA)*
*Faculty of Computer Science and Engineering, Office 431*
*Universidad Complutense de Madrid (UCM)*
*Calle Profesor José García Santesmases 9, Ciudad Universitaria, 28040 Madrid, Spain*

**Abstract**

Nowadays, mobile devices have become the natural substitute for the digital camera, as they capture everyday situations easily and quickly, encouraging users to express themselves through images and videos. These videos can be shared across different platforms exposing them to any kind of intentional manipulation by criminals who are aware of the weaknesses of forensic techniques to accuse an innocent person or exonerate a guilty person in a judicial process. Commonly, manufacturers do not comply 100% with the specifications of the standards for the creation of videos. Also, videos shared on social networks, and instant messaging applications go through filtering and compression processes to reduce their size, facilitate their transfer, and optimize storage on their platforms. The omission of specifications and results of transformations carried out by the platforms embed a features pattern in the multimedia container of the videos. These patterns make it possible to distinguish the brand of the device that generated the video, social network, and instant messaging application that was used for the transfer. Research in recent years has focused on the analysis of AVI containers and tiny video datasets. This work presents a novel technique to detect possible attacks against MP4, MOV, and 3GP format videos that affect their integrity and authenticity. The method is based on the analysis of the structure of video containers generated by mobile devices and their behaviour when shared through social networks, instant messaging applications, or manipulated by editing programs. The objectives of the proposal are to verify the integrity of videos, identify the source of acquisition and distinguish between original and manipulated videos.

## 1. Introduction

Mobile device manufacturers struggle every year to have the best features, benefits, functionality, and faster than the competition. People on a budget can buy a smartphone whose features solve most of their needs. The criteria that users value most when purchasing a mobile phone are design, performance, autonomy (battery), and integrated cameras, all of which are available in the low, medium, and high-end terminals.

Currently, some devices use artificial intelligence to manage up to 6 cameras, and the primary sensor that reaches 108 megapixels, which capture professional-quality photos and videos, but without the need to use a professional camera with heavy and expensive accessories.

According to the Global Mobile Trends 2020 report [1], the GSM Association predicts that by 2025 smartphone penetration will reach 80% globally, driven by India, Indonesia, Pakistan, Mexico, and a large number of users in Africa. Another feature that has burst more strongly during 2019 in the most advanced mobile ranges has been the 5G connectivity. According to Ericsson [2], this connectivity is expected to be present in up to 2,600 million subscriptions worldwide by 2025. They also say that the average monthly traffic per smartphone, which currently stands at 7.2 GB, could reach 24 GB by the end of 2025, which represents a significant increase in the amount of data that each mobile can move over a month. Some new user habits will be directly involved in this monthly gigabyte increase, which will promote greater use of the data daily. In fact, in these habits, the new virtual reality contents will have particular importance, which will be more and more popular, and therefore will consume more data from the terminals. While with 7.2 GB per month, we can enjoy HD video about 21 minutes a day, with 24 GB users will be able to play up to 30 minutes in HD with up to six extra minutes of virtual reality content, which will be a significant improvement in the consumption of higher quality content. On the other hand, social platforms (social networks, instant messaging applications) have become the most used communication channel by users as a source of information and to suggest, claim, consult or share personal experi-ences through multimedia files (photos and videos). According to [3], currently, Facebook is the social network with the most significant number of users in the world, with more than 2,414 million users, followed by Youtube com 2,000 million and Instagram with 1,000 million. In terms of messaging applications, WhatsApp has more than 1,600 million users worldwide, followed by Facebook Messenger with 1,300 million.

*Corresponding author
*Email addresses:* cquinto@ucm.es (Carlos Quinto Huama´n), asandoval@fdi.ucm.es (Ana Lucila Sandoval Orozco), javiergv@fdi.ucm.es (Luis Javier Garc´ıa Villalba)

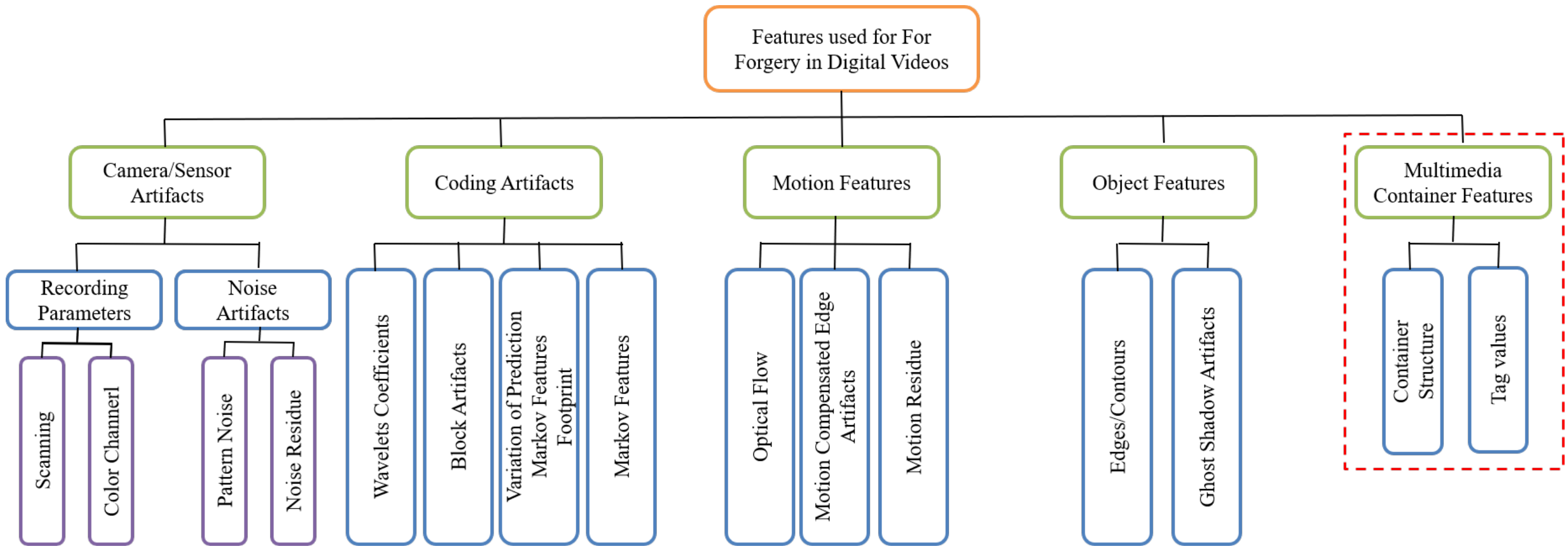

Figure 1: Features used to detect manipulation in digital videos.

In this context, cyber-attacks or privacy abuses and other forms of social and economic disruption increasingly affect citizens. Likewise, it is increasingly common for citizens, companies, and institutions of all kinds to carry out digital interactions and transactions in the social, financial and commercial fields, without being clear that the platform they use (Internet) also gives rise to cyber-crimes, cyber-attacks on critical infrastructures and violations of their privacy. To anticipate, prevent, and manage these threats, it is necessary to understand the causes, develop, and apply innovative technologies, solutions, forecasting tools, and knowledge. The videos captured by mobile devices and shared by social networks are also vulnerable to this type of attack, generating much interest in the development of intelligent software platforms that facilitate forensic investigations: In [4], a video analysis platform is developed to investigate criminal and terrorist activities, achieving efficient results that comply with legal, ethical, and privacy standards. In [5], they develop technologies to identify, categorize, and classify the source of images and video, as well as to verify the integrity of their content. This work is part of the RAMSES Project [6], which aims to design and develop a holistic, intelligent, scalable, and modular platform for Law Enforcement Agencies (LEAs) and to facilitate digital forensic investigations. To this, RAMSES brings together the latest technologies to develop an intelligent software platform, combining scraping of the public and deep web, detecting manipulation and steganalysis for images and videos, tracking malware payments, extraction and analysis of malware samples and Big Data analysis and visualizations tools. Our innovative method allows verifying the integrity and authenticity of digital videos using the intrinsic features of multimedia containers. Concretely, it can determine the brand of the mobile device, the social network or instant messaging application, and the editing program that was used for the manipulation. Additionally, it identifies anomalies in the values caused by steganography tools. This work is structured in 6 sections, the first being this introduction. Section 2 examines the related works in forensic video analysis and particularly in the study of container structures. In Section 3, the proposed analysis is carried out. Section 4 describes counter-forensic methods. Finally, section 5 presents the conclusions of the work.

## 2. Forensic Analysis for Digital Videos

Most investigations to detect tampering, identify source acquisition, integrity, and authenticity analysis have been performed for static photographic images [7] [8] [9] [10]. These techniques of forensic image analysis can be applied to the different previously extracted frames from a digital video. However, the scientific investigation requires solutions to the forensic issues related to video signals due to their peculiarities and the full range of possible alterations that can be applied.

In [11], the authors classify five methods of video manipulation detection: active techniques, passive techniques, intra-frame techniques, inter-frame techniques, detection techniques, and localization techniques. They also make a classification of the different types of features and methods used to detect the manipulation or falsification of a digital video. Among the types of features are sensor artifacts, coding artifacts, motion features, and object features. Among the types of techniques are inter-frame forgery detection and intra-frame forgery detection. However, it is necessary to add to this classification the features of the multimedia container, which allows the integrity and authenticity of digital videos to be accurately determined. Figure 1 presents an updated classification of the types of features used to detect tampering in digital video.

In this context, research has emerged in recent years that analyses the internal structure of multimedia containers, but mostly focuses on AVI formats, with minimal literature for MP4, MOV, and 3GP containers. Likewise, others analyze the structures of a small number of social networks and editing programs. It must be taken into account that mobile device manufacturers are continuously adding new atoms, labels, values, and other features related to each brand, model, operating system, and even the version of the latter. Social networks add different features depending on the versions of their applications and the device used to share the video. Similarly, the editing programs insert different structures depending on the type of manipulation and the version used.

Table 1: Main multimedia containers

| Container | Owner | Metadata | Frame rate | Bit rate | Video codec | Audio coded | OS |
|---|---|---|---|---|---|---|---|
| **MOV [12]** | Apple Inc. | Yes | Yes | Yes | H.264, H265 | AAC | iOS |
| **MP4 [13]** | Mpeg | Yes | Yes | Yes | H.264, H265 | AAC | Android |
| **3GP [14]** | 3GPP | Yes | Yes | Yes | H.264 | HE-AAC v2 | Android |
| **AVI [15]** | Microsoft | Yes | Yes | Yes | VFW/MJPEG | ACM/PCM | Windows |

In [16] [17], a study is made of the features that can be subject to forensic analysis on mobile devices. The biggest problem with this approach is that the different models of digital cameras use components from a small number of manufacturers and that the algorithms they use to generate the images and videos are also very similar between models of the same brand.

In [18], a detailed comparison of the main groups of acquisition source identification techniques. These are divided into five groups and are based on metadata, image features, CFA matrix defects, and colour interpolation, sensor imperfections, and wavelet transforms.

In [19], the authors perform a technique to verify the integrity of AVI format videos, generated by Video Event Data Recorders (VEDRs), an analysis of the structure of 296 original videos was performed, which were later edited by five editing programs. The results of the study showed that the editors significantly changed the structure and metadata values compared to the originals. Each editing program embeds a specific structure that helps the forensic analyst to detect whether a video has been manipulated in any way.

In [20], the authors analyzed the structures of AVI and MP4 format videos, grouped in 19 models of digital cameras, 14 models of mobile phones and six editing programs. After analyzing the original videos, the authors determined that the structures of each type of container are not strictly defined as specified in the standards. Considerable differences were found between videos generated by such devices. Also, the AVI videos, after being manipulated by the editing programs, changed the internal structure, including the metadata values, essential features to know the origin of the videos.

In [21], the authors implemented an unsupervised method for verifying video integrity based on the dissimilarity between original and edited videos. They also developed a technique to identify the source of video acquisition by analyzing containers. To achieve this goal, they used the *MP4Parser* library, obtaining Extensible Markup Language (XML) files for later analysis, making great results in their experiments, arguing that the solution uses a minimum of computational resources compared to other alternatives.

### *2.1. Digital Video*

Digital videos are composed of a sequence of images and audio that are previously encoded and then encapsulated in a multimedia container [22]. For the generation of digital videos, two procedures are performed in parallel (image sequence processing and audio sequence processing) [23]. On the one hand, the lens system captures the light of the scene by controlling exposure, focus, and image stabilization. This light enters the camera through the lens system and applies various filters to improve the visual quality of the image (infrared and anti-aliasing). The light then passes to the image sensor through the color filter array, which are light-sensitive elements called pixels. This signal is converted into a digital signal, which is transmitted to the Digital Signal Processor (DSP) and specifically to the Digital Image Processor (DIP), which performs different camera processes to stabilize the signal and correct artifacts such as removing noise and other introduced anomalies [24] [25]. Then the compression process is carried out using a codec, after synchronization and finally, it is encapsulated in a multimedia container.

On the other hand, the sound signal, transmitted through the air, is captured through the microphone that works as an electro-acoustic sensor. The microphone transforms the sound waves into an electrical signal to increase its intensity and transmit it to an Analog to Digital Converter (ADC), which converts it into a digital signal. The Digital Audio Processor then improves the quality of the audio before compression (volume and frequency control). Then the signal is compressed with an encoding algorithm and finally encapsulated in a multimedia container.

In [26], it is indicated that video compression is achieved by eliminating the temporal redundancy that exists between the frames that make up a sequence, thus only obtaining components necessary for faithful reproduction of the data. There are different video compression standards, but currently, the most used by mobile devices are H264/AVC or MPEG-4 Part 10 [27] and H265/HEVC or MPEG-H Part 2 [28].

### *2.2. Multimedia Containers*

A wide variety of formats, called media containers, are used to store video, audio, metadata, sync information, and error correction information, following a pre-defined format in current standards. As its name suggests, a multimedia container is a file that contains several elements, at least the video and audio tracks [29] [30] [12]. Some containers also allow you to include other elements such as images or subtitles without the need for external files. Figure 2 shows the elements of the media container and the interlacing of the content. The multimedia containers most used by mobile devices currently are MP4, MOV, and 3GP. Table 1 presents the primary multimedia containers with their respective descriptions.

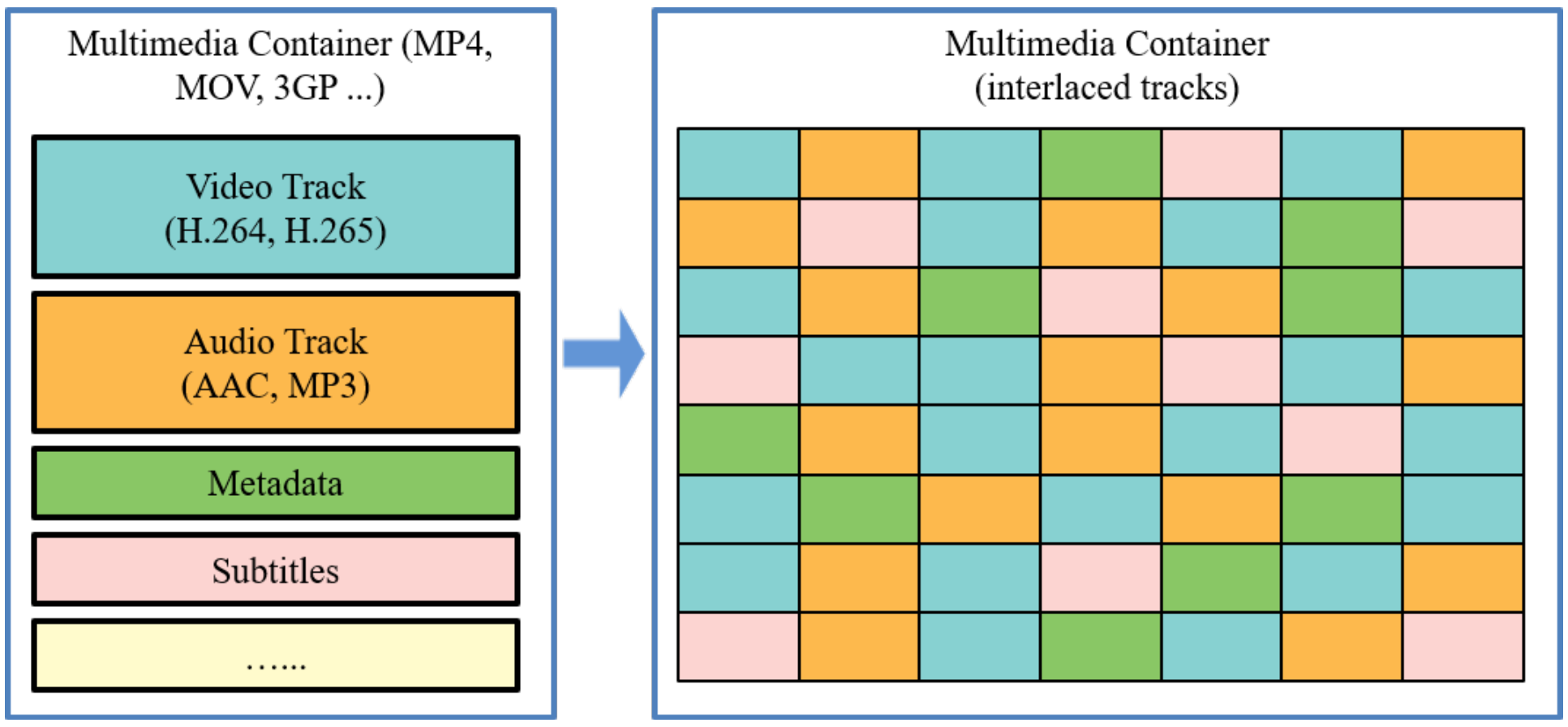

Figure 2: Multimedia container elements and content interlacing

Metadata records information related to the conditions of image/video capture, such as date and time of generation, presence or absence of flash, the distance of objects, exposure time, shutter aperture and GPS information, among others. In other words, information of interest that complements the main content of a digital document. Metadata is also used as a powerful help to organise the search within image and video libraries.

Digital images are stored in a variety of formats such as TIFF, JPEG [31] or PSD. Some metadata containers for the different image formats are IFD Exif, TIFF, Adobe XMP, and IPTC-IIM. The Exif specification [32] is most commonly used as the most common metadata container in digital cameras [33]. The Exif specification includes hundreds of labels, including brand and model, although it is noteworthy that the specification itself does not make its existence mandatory in archives.

In the videos, the metadata is stored in the media container and key-value format. They are also distributed in all the tracks of the video.

### 2.3. *MP4, MOV, and 3GP Containers*

Currently, 100% of mobile device manufacturers use MP4, MOV, or 3GP multimedia containers to encapsulate digital video captured by integrated cameras. All three containers are based on the MPEG-4 part-14 [13] and QuickTime [12] standards, which define their structure and content. In this context, the MP4, MOV, and 3GP multimedia containers are composed of a set of elements called atoms. A 32-bit unsigned integer specifies the types of atoms, typically interpreted as a four-character lowercase ASCII code. The atoms are hierarchical by nature. That is, an atom can contain other atoms, which in turn can contain other atoms, and so on. Each atom carries its size and type information as well as its data.

The format of the data stored in a given atom can not always be determined only by the "type" field of the atom. The type of the parent atom can also be important. In other words, a given type of atom may contain different types of information depending on its root atom. Atoms usually do not follow any particular order. The head of the atom contains the following fields:

- **size**: A 32-bit unsigned integer that indicates the size of the atom. However, this field may contain special values indicating an alternative method for determining the size of the atom. These special values are normally used only for data medium "mdat" atoms:
  - **0**: Allowed only by a higher level atom, designates the last atom in the file and indicates that the atom extends to the end of the file
  - **1**: Indicates that the actual size of the atom is in the "extended size" field.

  The actual size of an atom can not be less than 8 bytes.
- **type**: A 32-bit integer that specifies the type of atom. For example, "moov" 0x6D6F6F76 for a film atom, "trak" 0x7472616B for a track atom. Knowing the type of an atom allows you to interpret your data.
- **extended size**: A 64-bit field that is used by atoms with data containing more than 232 bytes. In this case the "size" field is set to 1.

Some atoms also contain the "version" and "flags" tags. These atoms are called complete atoms and are not treated as part of the atom header but as specific data fields for each type of atom containing them. The main types of atoms are:

- **"ftyp"**: File compatibility type, identifies the file type and differentiates it from similar file types, such as MPEG-4 and JPEG-2000 files.
- **"mdat"**: Media sample samples from films such as frames and groups of video audio samples. Usually, this data can be interpreted only by using the movie resource.
- **"free"**: The unused space available in the file.
- **"moov"**: Metadata of movie resources (number and type of tracks, location of sample data, and so on). Describe where they are and how the movie data is interpreted. This atom contains at least two "trak" atoms, one to store the video track information, and one for the audio track. Both atoms have a similar structure. However, the most recent mobile devices generate videos that contain more than two "trak" atoms and add new child atoms.

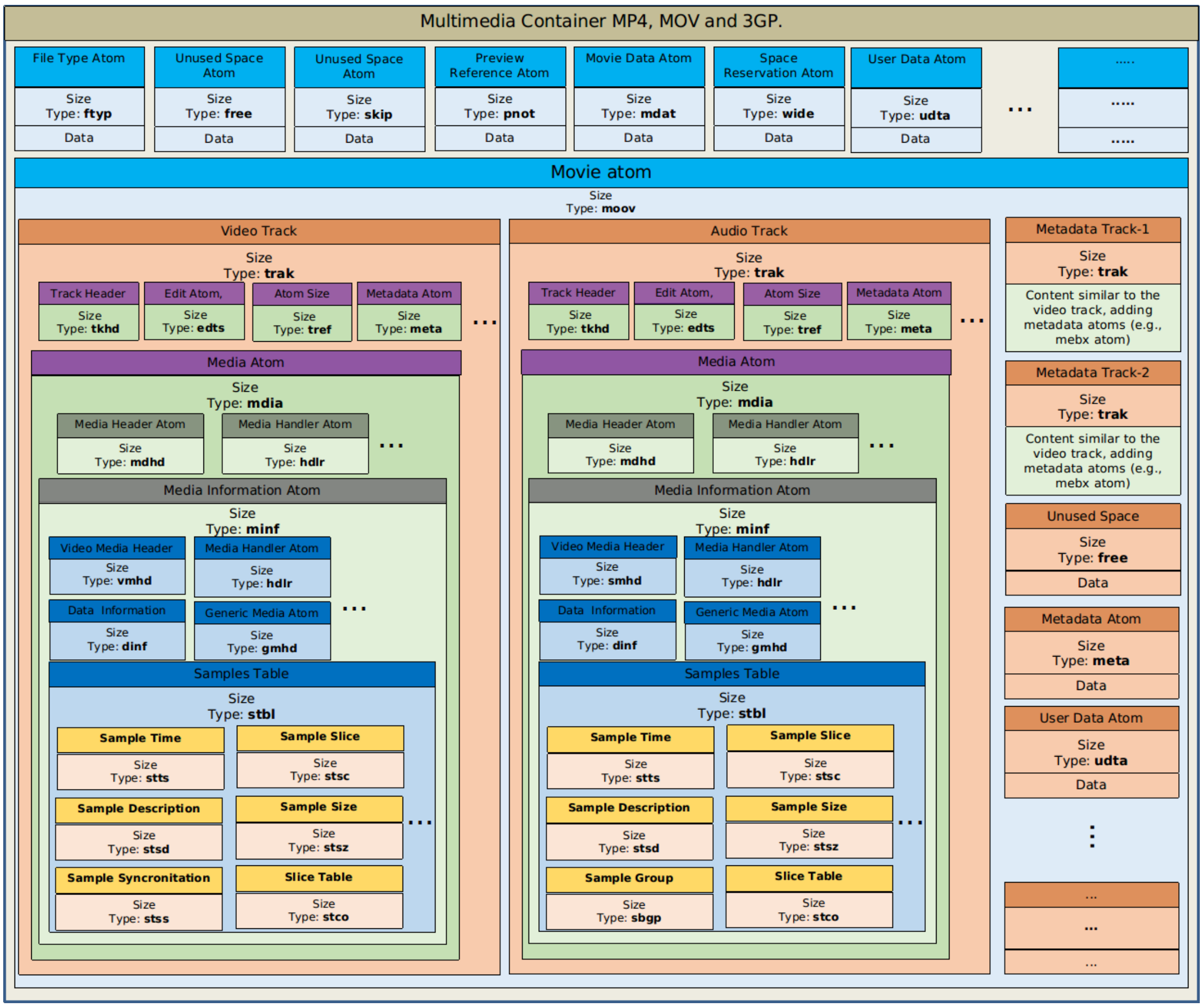

Figure 3: Structure of the MP4, MOV, and 3GP multimedia containers

Table 2: Atoms format

| Atom | Contain | |
|---|---|---|
| **"ftyp"** | *Size* | Specifies the number of bytes in this atom (32 bits). |
| | *Type* | "ftyp" (32 bits); |
| | *Major Brand* | Identifies the type of movie file. If a file is compatible with multiple marks, these are specified in the "Compatible Brands" fields, and in this field the preferred mark (32 bits) is identified. |
| | *Minor version* | Identifies the type of movie file, it is represented in binary coded decimal form (BCD) indicating year, month and a code in binary zero. For example "BCD 20 04 06 00" (32 bits). |
| | *Compatible Brands* | List of supported file formats (32 bits). |
| **"mdat"** | *Size* | Specifies the number of bytes in this atom (32 bits). |
| | *Type* | "mdat" (32 bits). |
| | *Data* | Contains the audio and video data of the movie. |
| **"free"** | *Size* | Specifies the number of bytes in this atom (32 bits). |
| | *Type* | "free" (32 bits). |
| | *Free Space* | Contains bytes of free space. These bytes are all 0 (32 bits). |
| **"moov"** | *Size* | Indicates the number of bytes in this atom (32 bits). |
| | *Type* | "moov" (32 bits). |

These atoms contain the fields shown in Table 2.

Figure 3 shows the updated structure of the MP4, MOV, and 3GP multimedia containers.

Therefore, in the videos, the atoms are features that distinguish each brand, model, social platform and editing program. Hierarchically, an atom can be represented as follows: /moov/. In turn, these atoms have child atoms, e.g., *Path-tag: /moov/mvhd/*. Each atom can have a set of tags, e.g., */moov/mvhd/ version*. Also, these tags contain values e.g., *Path-tag-value: /moov/mvhd/ version,* **value***: 0*. This feature set is called the multimedia container structure.

## 3. Multimedia Container Structure Analysis

This work proposes a technique for the verification of video integrity, identification of the source of acquisition, and differentiation between original and manipulated videos. Three phases compose the technique: 1) Preparation of the set of original and manipulated videos; 2) Extraction of the information contained in each video with the atom extraction algorithm; 3) Multimedia containers structure analysis. Figure 4 shows the three phases for the multimedia containers structure analysis.

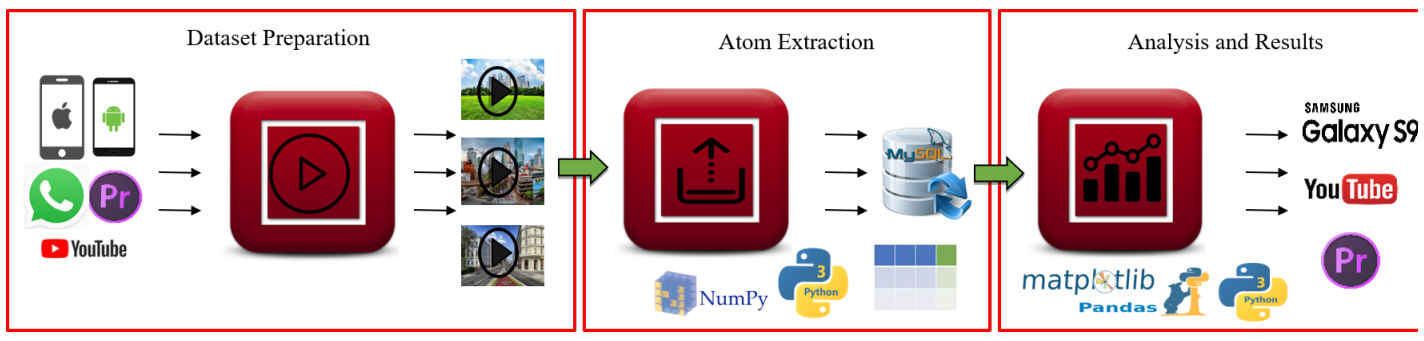

Figure 4: Phases of the proposed technique

### 3.1. Dataset Preparation

In the literature, there are few video datasets generated by mobile devices that have been shared by Social Networks (SN), Instant Messaging Applications (IMA), and Editing Programs (EP). Having an organized, robust, balanced, and above all, up-to-date video data set is a challenge. For that reason, a heterogeneous and sufficiently large dataset has been generated, to evaluate the proposed method in real scenarios and thus obtain valid results. Tables 3 and 4 show the set of 1957 original videos generated by 86 devices, 66 models, and 11 brands of mobile devices most used today. These videos were generated with the default settings, different camera angles, different light conditions, and scenes.

Tables 5 and 6 show the set of videos shared through SN and IMA. For this purpose, 100% of the videos from the devices marked with (*) in Table 3 were shared, making a total of 3682 videos. Also, the settings that were used to perform the process of uploading, downloading, and transfer of videos are detailed. In Table 7, the 721 videos manipulated with five EP and their respective configurations are presented. For this purpose, 100% of videos from the devices marked with (*) in Table 3 were used.

### 3.2. Atom Extraction

Atom extraction consists of storing the atoms, tags, values, order of appearance of atoms, and all kinds of relevant information from the video generated by a mobile device. To process this information, we use the Atom Extraction Algorithm [34]. The process starts by obtaining the initial byte, size, and type of atom with a length of 4 bytes, represented by a string of characters. Next, the duplicity of atoms and the existence of child atoms are verified. Finally, a DataFrame is obtained with the set of atoms and tags with their respective values and order of appearance.

### 3.3. Multimedia Containers Structure Analysis

The features that come from the extraction of atoms are structured in a DataFrame with the following columns: *Path-file-name* is the path of the video followed by the video name, *Path-origin* is the path of the video, *Class-label* is the identification of each social network and messaging application. *File-name* is the name of the video, *Marker* is the brand of the mobile device, *Model* is the model of the mobile device, *PathOrder-tag* is the set of atoms with their relative order, *Value* is the value of each label, *Reading-orders* is the absolute order of each atom, also called the order of appearance. Table 8 details the contents of the DataFrame.

Given a video $X$, it contains a set of atoms ($a_1$,. $a_n$), represented by */ftyp-1/*. These atoms may contain other atoms and tags: $a=((a_1), w_1$ $(a_n), w_n)$ represented by */ftyp-1/majorBrands*. Also, these tags have values: */ftyp-1/majorBrands: @mp42*. In this sense, the *PathOrder-tag* includes 2 types of features: Sequence of atoms *(/ftyp-1/)* and sequence of atoms with their tags *(/ftyp-1/majorBrands)*.

To know the structure that each tool inserts into the multimedia containers, we compared the structure of an original video (Apple brand, Ipad Air model), with the same video after being shared through Facebook HD, Youtube, WhatsApp, Linkedin, Telegram. Figures 5 and 6 show the individual comparison of the multimedia container structure.

We can see that the original video has four root atoms *(ftyp-1, wide-2, mdat-3, moov-4)*. The value of the *@majorBrands* and *compatibleBrand* tags is *qt* in both cases. The path to get the value of the *@timeScale* tag is */ftyp-1/wide-2/mdat-3/moov-4/mvhd-1/@timeScale:600*. The *moov-4* atom contains four *trak* atoms, 1 for the video track, 1 for the audio track, and two exclusively for metadata. The structure of the container is composed of 651 features.

The Facebook HD video, it has four root atoms *(ftyp-1, moov-2, free-3, mdat-4)*. The values of the @compatibleBrand and @majorBrands tags are *isomiso2avc1mp41* and *isom* respectively. The path to get the value of the *@timeScale* tag is */ftyp-1/moov-2/mvhd-1/@timeScale:1000*. The *moov-2* atom contains two *trak* atoms, 1 for the video track, 1 for the audio track. The structure of the container is composed of 267 features.

The Youtube video, it has three root atoms *(ftyp-1, moov-2, mdat-3)*. The values of the @majorBrands and @compatibleBrand tags are *mp42* and *isommp42* respectively. The path to get the value of the *@timeScale* tag is */ftyp-1/moov-2/mvhd-1/@timeScale:1000*. The *moov-2* atom contains two *trak* atoms, 1 for the video track, 1 for the audio track. The structure of the container is composed of 250 features.

The WhatsApp video, it has four root atoms *(ftyp-1, beam-2, moov-3, mdat-4)*. The values of the @majorBrands and @compatibleBrand tags are *mp42* and *mp41mp42isom* respectively. The path to get the value of the *@timeScale* tag is /ftyp-1/beam-2/moov-3/mvhd-1/@timeScale:44100. The *moov-3* atom contains two *trak* atoms, 1 for the video track, 1 for the audio track. The structure of the container is composed of 237 features.

Table 3: Original video dataset

| Brand (ID) | Model (ID) | Id Device | OS Version | Video Codec | #vOrig | SN and MSN Apps | Editing Programs |
|---|---|---|---|---|---|---|---|
| APPLE (B01) | Ipad 2 (M01) | D01 | iOS 7.1.1 | H.264 | 16 | | |
| | | D02 | iOS 9.3.5 | H.264 | 20 | x | x |
| | Ipad Air (M02) | D03 | iOS 11.3 | H.264 | 20 | x | x |
| | Ipad Mini (M03) | D04 | iOS 8.4 | H.264 | 16 | | |
| | Iphone 4 (M04) | D05 | iOS 7.1.2 | H.264 | 19 | | |
| | Iphone 4S (M05) | D06 | iOS 7.1.2 | H.264 | 20 | | |
| | | D07 | iOS 8.4.1 | H.264 | 15 | | |
| | Iphone 5 (M06) | D08 | iOS 7.0.4 | H.264 | 20 | x | x |
| | | D09 | iOS 9.3.3 | H.264 | 19 | | |
| | | D10 | iOS 8.4 | H.264 | 32 | | |
| | Iphone 5C (M07) | D11 | iOS 7.0.3 | H.264 | 19 | | |
| | | D12 | iOS 8.4.1 | H.264 | 21 | | |
| | | D13 | iOS 10.2.1 | H.264 | 19 | | |
| | Iphone 5S (M08) | D14 | iOS 9.2 | H.264 | 20 | x | |
| | Iphone 6 (M09) | D15 | iOS 8.4 | H.264 | 17 | x | x |
| | | D16 | iOS 9.2 | H.264 | 20 | | |
| | | D17 | iOS 10.1.1 | H.264 | 18 | | |
| | | D18 | iOS 11.2.0 | H.264 | 20 | | |
| | Iphone 7 (M10) | D19 | iOS 11.2.6 | H.264 | 24 | x | x |
| | | D20 | iOS 10.2.6 | H.264 | 20 | | |
| | | D21 | iOS 11.0.3 | H.264 | 20 | | |
| | Iphone 7 Plus (M11) | D22 | iOS 11.0.3 | H.264 | 20 | | |
| | Iphone 8 (M12) | D23 | iOS 12.1.4 | H.264 | 18 | | |
| | Iphone 8 Plus (M13) | D24 | iOS 11.2.5 | H.265 | 18 | x | x |
| | | D25 | iOS 11.2.6 | H.265 | 21 | | |
| | Iphone 6 Plus (M14) | D26 | iOS 10.2.1 | H.264 | 19 | | |
| | Iphone XR (M15) | D27 | iOS 12.1.1 | H.264 | 16 | | |
| | | D28 | iOS 12.1.4 | H.265 | 16 | | |
| | Iphone X (M16) | D29 | iOS 11.4.1 | H.264 | 20 | x | |
| | Iphone XS (M17) | D30 | iOS 12.1.1 | H.264 | 20 | | |
| | Iphone XS Max (M18) | D31 | iOS 12.1.0 | H.264 | 20 | x | |
| BQ (B02) | Aquaris E5 (M19) | D32 | Android | H.264 | 40 | | |
| | Aquaris E4.5 (M20) | D33 | Android | H.264 | 40 | | |
| Huawei (B03) | Ascend (M21) | D34 | Android | H.264 | 19 | x | x |
| | P8 (M22) | D35 | Android | H.264 | 19 | | |
| | | D36 | Android | H.264 | 20 | | |
| | P9 (M23) | D37 | Android | H.264 | 19 | x | x |
| | P9 Lite (M24) | D38 | Android | H.264 | 19 | | |
| | Y5 (M25) | D39 | Android | H.264 | 20 | | |
| | Y635-L01 (M26) | D40 | Android | H.264 | 20 | | |
| | | D41 | Android | H.264 | 21 | | |
| | Honor 5C (M27) | D42 | Android | H.264 | 19 | | |
| | P10 (M28) | D43 | Android | H.264 | 20 | x | |
| | P30 (M29) | D44 | Android | H.264 | 20 | | |
| | PSmart Plus (M30) | D45 | Android | H.264 | 20 | | |
| | G7 (M31) | D46 | Android | H.264 | 20 | | |

Table 4: Original video dataset

| Brand (ID) | Model (ID) | Id Device | OS Version | Video Codec | #vOrig | SN and MSN Apps | Editing Programs |
|---|---|---|---|---|---|---|---|
| LG (B04) | D290 (M32) | D47 | Android | H.264 | 19 | | |
| | Nexus 5 (M33) | D48 | Android | H.264 | 30 | x | |
| | | D49 | Android | H.264 | 26 | | |
| | G6 (M34) | D50 | Android | H.264 | 30 | x | |
| Microsoft (B05) | Lumia 640 LTE (M35) | D51 | Windows Phone | H.264 | 80 | x | |
| Motorola (B06) | Moto G1 (M36) | D52 | Android | H.264 | 15 | | |
| | | D53 | Android | H.264 | 15 | | |
| | Moto G2 (M37) | D54 | Android | H.264 | 15 | x | |
| | Nexus 6 (M38) | D55 | Android | H.264 | 15 | x | |
| | G7 Play (M39) | D56 | Android | H.264 | 20 | | |
| Nokia (B07) | 808 Pureview (M40) | D57 | Symbian | H.264 | 80 | | |
| One Plus (B08) | A0001 (M41) | D58 | Android | H.264 | 20 | x | |
| | A3000 (M42) | D59 | Android | H.264 | 19 | | |
| | A3003 (M43) | D60 | Android | H.264 | 36 | | |
| Samsung (B9) | Galaxy A6 (M44) | D61 | Android | H.264 | 20 | x | x |
| | Galaxy Nexus (M45) | D62 | Android | H.264 | 20 | | |
| | Galaxy GT-I8190N (M46) | D63 | Android | H.264 | 22 | | |
| | Galaxy GT-I8190 (M47) | D64 | Android | H.264 | 16 | | |
| | Galaxy S3 (M48) | D65 | Android | H.264 | 20 | | |
| | Galaxy GT-I9300 (M49) | D66 | Android | H.264 | 19 | | |
| | Galaxy S3 Neo (M50) | D67 | Android | H.264 | 18 | | |
| | Galaxy S4 Mini (M51) | D68 | Android | H.264 | 39 | | |
| | Galaxy S4 (M52) | D70 | Android | H.264 | 20 | | |
| | Galaxy S5 (M53) | D71 | Android | H.264 | 20 | | |
| | | D72 | Android | H.264 | 19 | x | |
| | Galaxy S6 (M54) | D73 | Android | H.264 | 20 | | |
| | Galaxy S7 (M55) | D74 | Android | H.264 | 20 | x | |
| | Galaxy S9 Plus M56 | D75 | Android | H.264 | 20 | | |
| | | D76 | Android | H.265 | 20 | x | x |
| | Galaxy J5 2016 (M57) | D77 | Android | H.264 | 24 | x | x |
| | Galaxy Tab 3 (M58) | D78 | Android | H.264 | 37 | | |
| | Galaxy Trend Plus (M59) | D79 | Android | H.264 | 16 | | |
| | Galaxy Tab A (M60) | D80 | Android | H.264 | 16 | x | x |
| Wiko (B10) | Ridge 4G (M61) | D81 | Android | H.264 | 88 | | |
| Xiaomi (B11) | Mi3 (M62) | D82 | Android | H.264 | 24 | x | |
| | Redmi Note 3 (M63) | D83 | Android | H.264 | 20 | | |
| | Redmi Note 5 (M64) | D84 | Android | H.264 | 20 | x | x |
| | PocoPhone (M65) | D85 | Android | H.264 | 20 | x | |
| | Redmi 5 Plus (M66) | D86 | Android | H.264 | 20 | | |
| **Total videos** | | | | | **1957** | | |

Table 5: Video dataset (Social networks)

| Social Network | Version | Upload Process | Download Process | Edited Container | Shared Videos |
|---|---|---|---|---|---|
| Facebook HD (SN01) | Website | Max 4gb, 240min | Firefox (Inspect element) | MP4 | 310 |
| Facebook SD (SN02) | Website | Max 4gb, 240min | Firefox (Inspect element) | MP4 | 290 |
| Youtube (SN03) | Website | Max 128gb,12hrs | Youtube studio beta | MP4 | 310 |
| Flickr (SN04) | Website | Max 1gb | Website (save as) | MP4 | 508 |
| Linkedin (SN05) | Website | Max 6gb, min 75kb | Website (save as) | MP4 | 310 |
| Instagram (SN06) | Website | Max 10 min, ratio 9:16 | Firefox (Inspect element) | MP4 | 310 |
| Twitter (SN07) | Website | Max 500mb, 2.20 min | Firefox(twitervideodownloader) | MP4 | 240 |
| Tumblr (SN08) | Website | Max 100mb | Firefox (Inspect element) | MP4 | 320 |
| **Total videos** | | | | | **2598** |

Table 6: Video dataset (Instant messaging applications)

| Messaging Application | Version | Features | Edited Container | Shared Videos |
|---|---|---|---|---|
| Facebook Msn (MA01) | 255.0.0.13.113 | - | MP4 | 270 |
| WhatsApp (MA02) | 2.19.20 | - | MP4 | 504 |
| Telegram (MA03) | 5.7.1 | Max 1.5GB | MP4 | 310 |
| **Total videos** | | | | **1084** |

Table 7: Video dataset (Editing Programs)

| Editing Program | Version | Settings | Container | Cantidad |
|---|---|---|---|---|
| Adobe Premiere (ES01) | 2018-12.0 | Open/Save | MP4 | 178 |
| Camtasia (ES02) | 2018.0.1 | Open/Save | MP4 | 177 |
| FFmpeg (ES03) | 3.4.4 | Copy -vcodec | MOV | 124 |
| Lightworks (ES04) | 14.5 | Import/Export | MP4 | 172 |
| Movie Maker (ES05) | 2012 | Open/Save | MP4 | 70 |
| **Total videos** | | | | **721** |

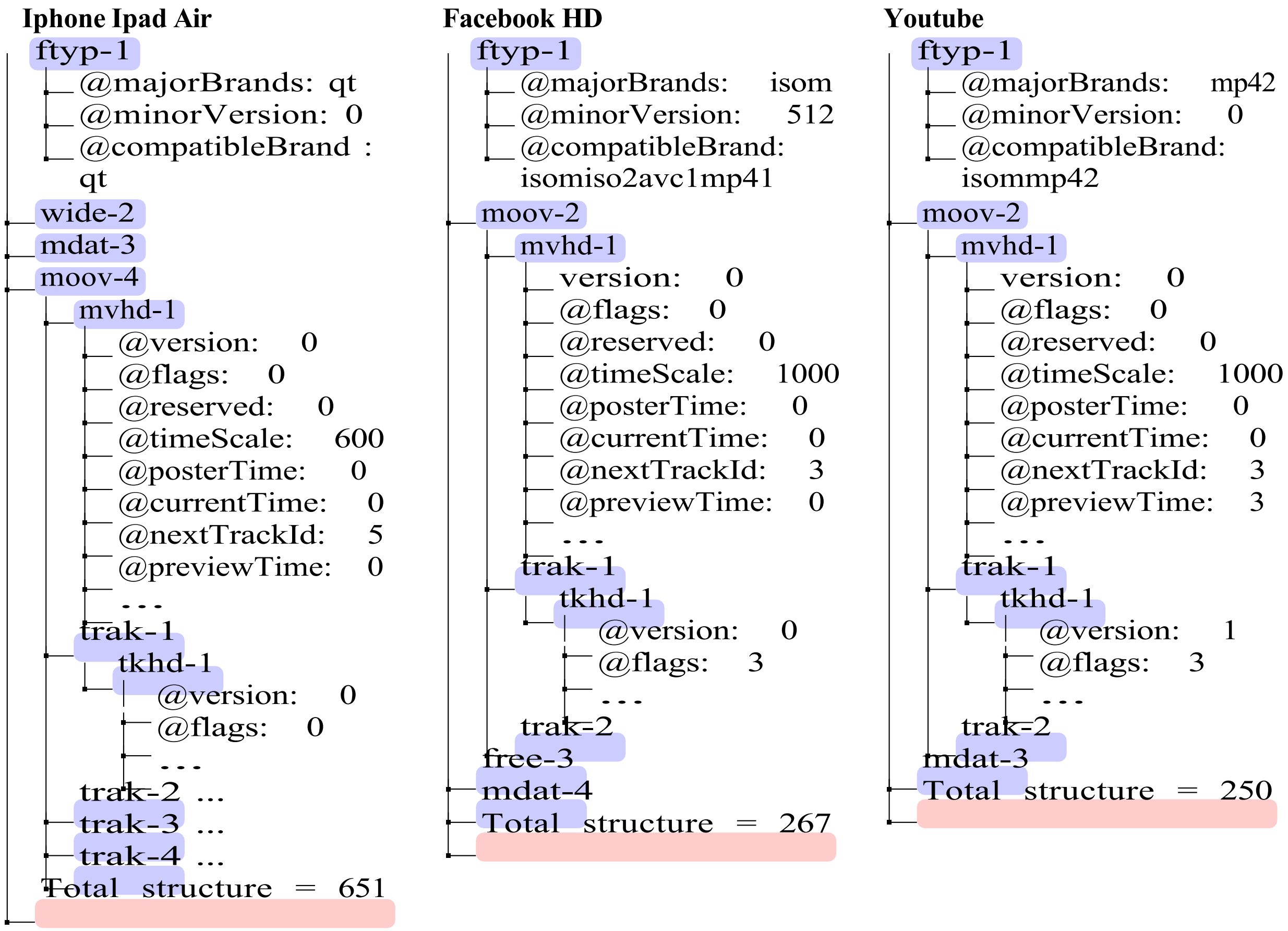

Figure 5: Comparison of the multimedia container structure

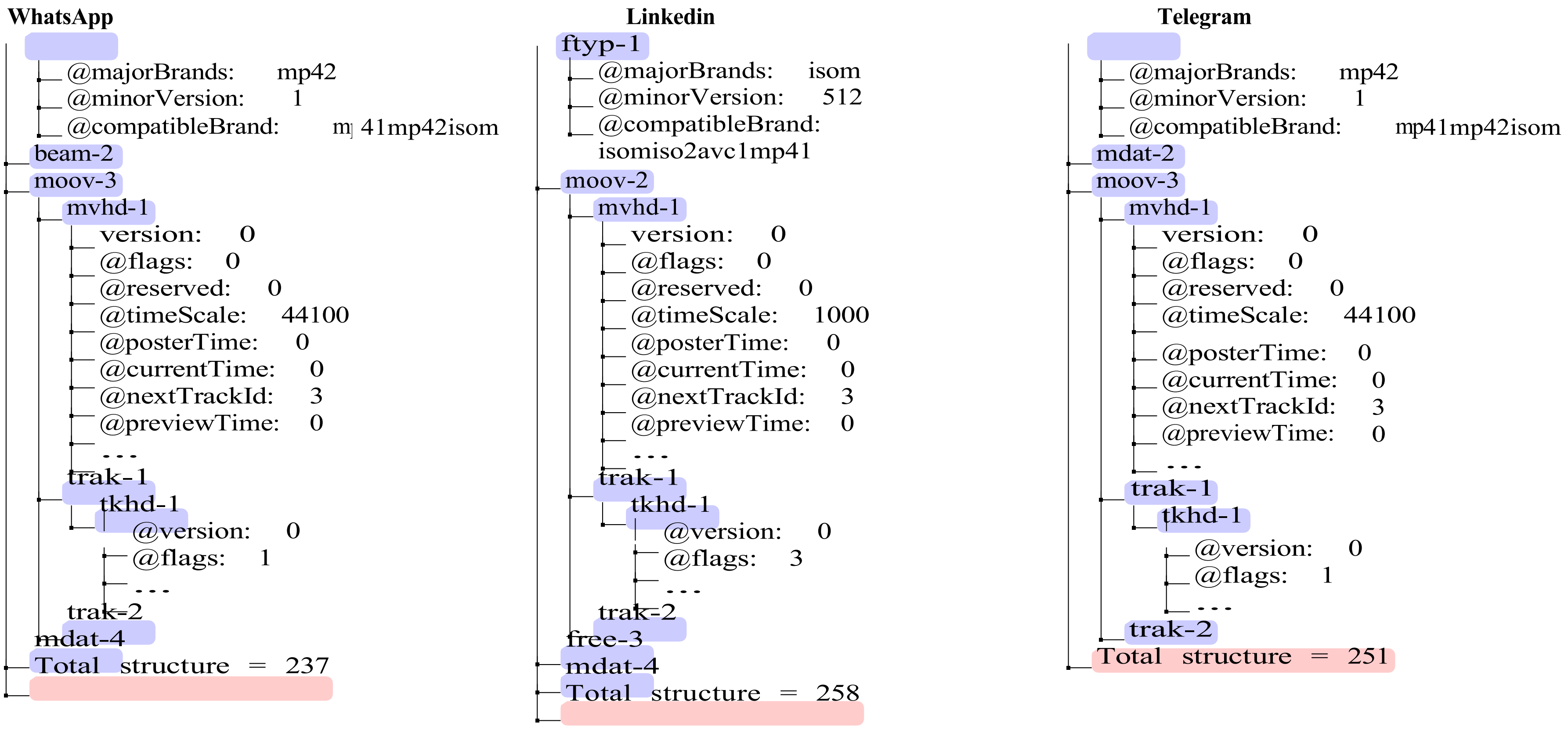

Figure 6: Comparison of the multimedia container structure

The Linkedin video, it has four root atoms *(ftyp-1, moov-2, free-3, mdat-4)*. The values of the @majorBrands and @compatibleBrand tags are *isom* and *isomiso2avc1mp41* respectively. The path to get the value of the *@timeScale* tag is */ftyp-1/moov-2/mvhd-1/@timeScale:1000*. The *moov-2* atom contains two *trak* atoms, 1 for the video track, 1 for the audio track. The structure of the container is composed of 258 features.

The Telegram video, it has three root atoms *(ftyp-1, mdat-2, moov-3)*. The values of the @majorBrands and @compatibleBrand tags are *mp42* and *mp41mp42isom* respectively. The path to get the value of the *@timeScale* tag is */ftyp-1/mdat-2/moov-3/mvhd-1/@timeScale:44100*. The *moov-3* atom contains two *trak* atoms, 1 for the video track, 1 for the audio track. The structure of the container is composed of 251 features. It should be noted that after analyzing the videos on the Flickr social network, that the platform does not make any re-compression and does not change the multimedia container structure, maintaining the same features as the original video.

In summary, this analysis has shown that each social network and instant messaging application inserts a different structure. These differences derive from the order of appearance of the atoms and values assigned to each tag.

### *3.3.1. Massive Structure Analysis*

To perform the massive analysis, we take advantage of the presence and absence of the extracted features. For this, binary variables (0 or 1) are used, assigning the value "1" for each *PathOrder-tag* that is present and "0" for the absent one, obtaining structures with equal cardinality. The t-Distributed Stochastic Neighbor Embedding (t-SNE), is a dimension reduction and visualization technique for high-dimensional data [35], [36]. We explore the applicability of t-SNE to the proposed datasets and make some observations. Figure 7 shows the distribution of the samples of each brand in two dimensions. The samples from Apple, Wiko, Nokia, BQ, Samsung, LG and Microsoft are well grouped and do not overlap between classes. However, Motorola and Xiaomi overlap each other, which means they have similar features. Figure 8 shows the distribution of samples from each social network. Although indeed, SN are not 100% grouped, its classification has a better prognosis. The samples from each editing program are better grouped, and none of them overlaps with another class (Figure 9).

Also, we use Principal Component Analysis (PCA) because the input dimension is high, and it allows us to visualize the data in two or three dimensions [37]. Scaling is not applied because the variables only have values "0" and "1". The original videos (1957) consolidate a total of 804 *PathOrder-tag* (804 dimensions), and by applying PCA with the configuration of the number of components parameter to 0.99 (capture the minimum number of principal components that retain 99% of the variances of the variables), we manage to reduce the number of dimensions to 20, and with a configuration of 0.95, we obtain 13 dimensions. The first principal component */wide-2/* contains 31.1% of the variance, the second */mdat-2/* contains 13.6% and the third */moov-2/meta-6/keys-2/* has 12.7%, adding up to 57.4%; while the remaining 17 components account for 41.6%

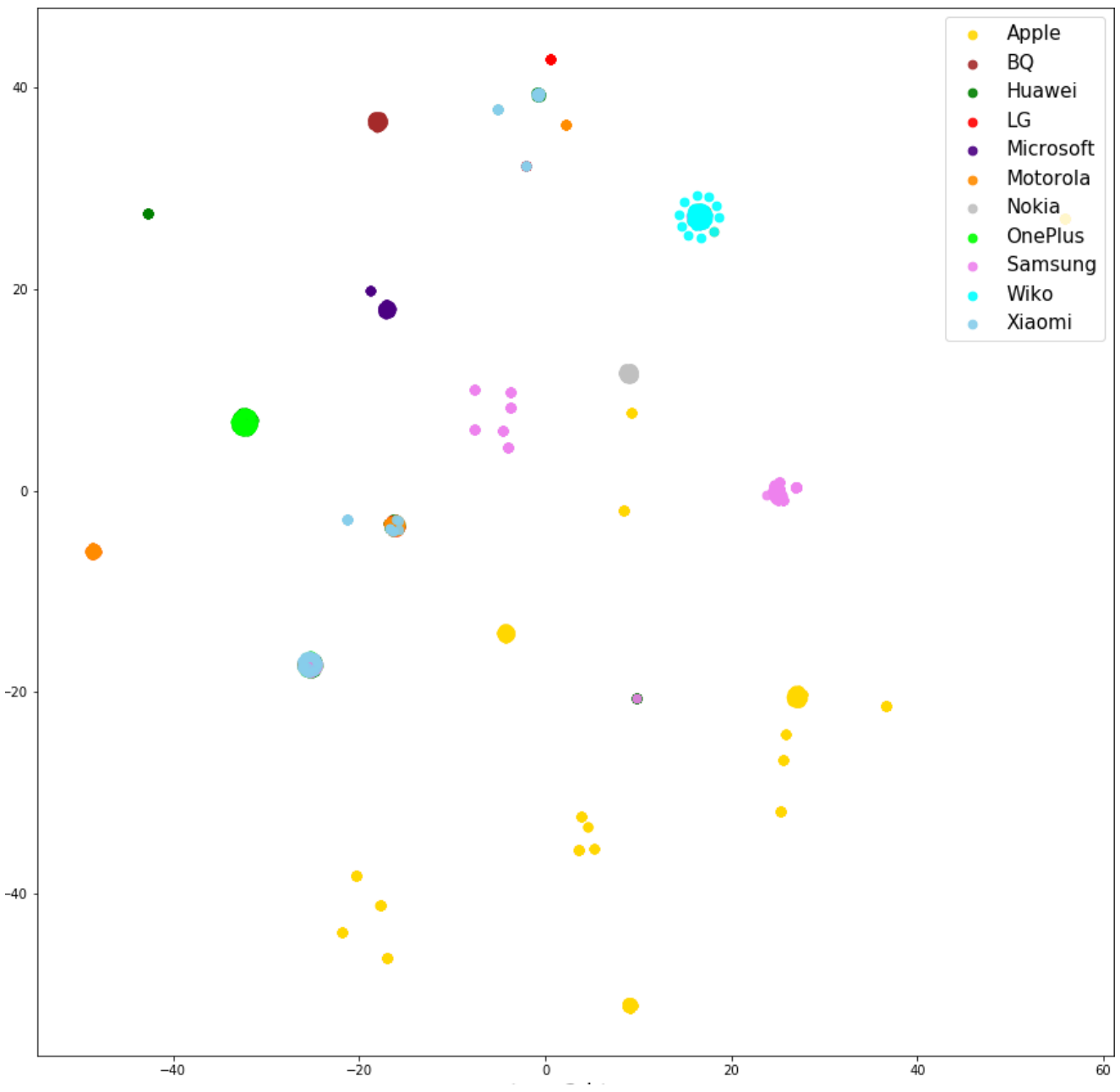

Figure 7: Visualization of the dataset (Original videos) using the t-SNE

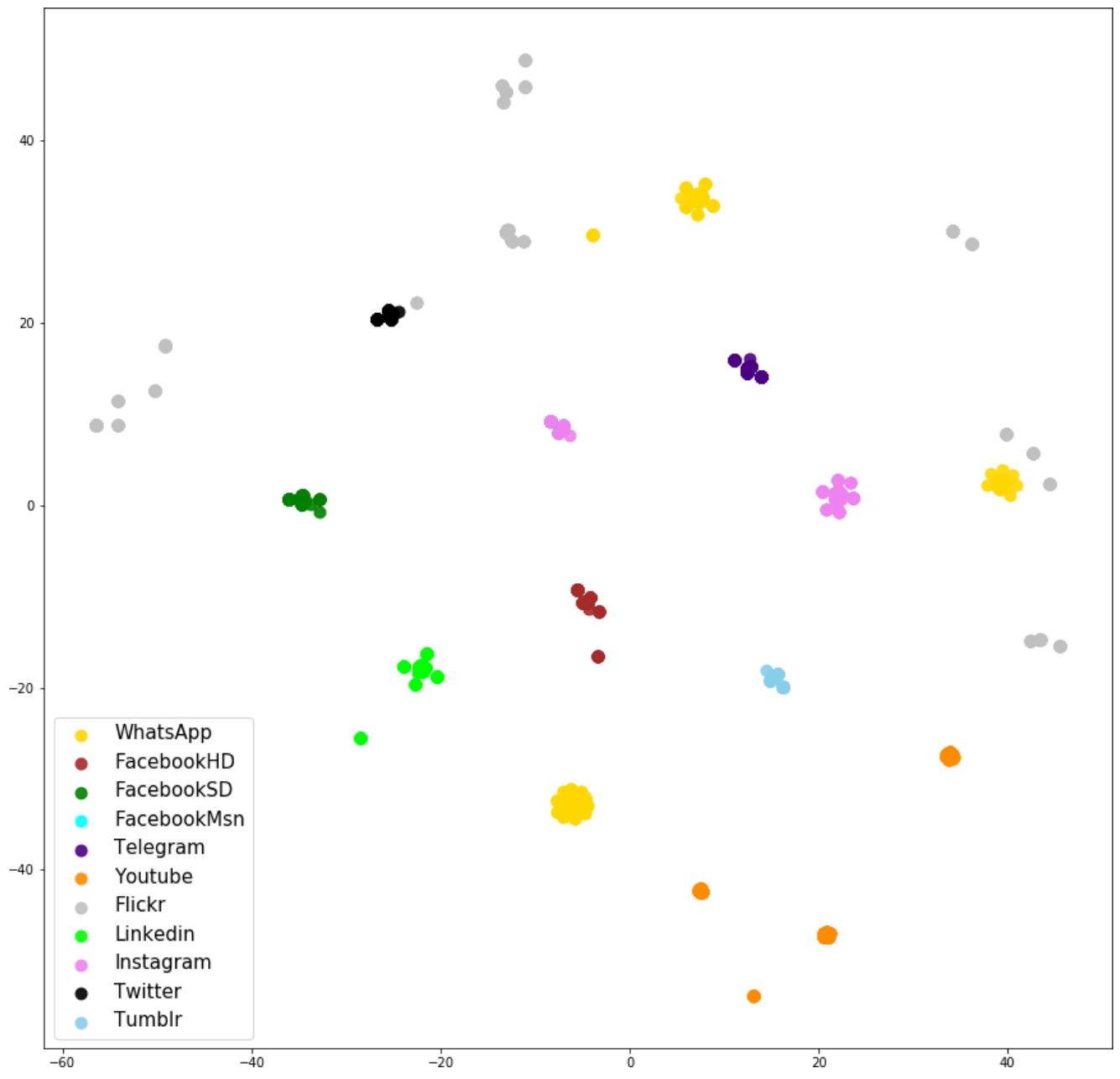

Figure 8: Visualization of the dataset (SN and IMA) using the t-SNE

of the variance. In Figure 10, the three-dimensional scatter plot is presented using the dataset of the brands, where it is seen that the videos of the Apple brand have *PathOrder-tag* similar and are grouped remarkably. However, Xiaomi, Huawei, and LG overlap because they have very similar features. In general, the differences between brands are considerable, and these differences can be improved by adding more mobile device models. On the other hand, the videos of the SN and IMA (3682) group a total of 842 dimensions, and applying PCA with a configuration of 0.99 and 0.95 was reduced to 24 and 15 dimensions, respectively. The first principal component */moov-*

Table 8: DataFrame content description

| Column | Description | Example |
|---|---|---|
| *Path-file-name* | Contains the absolute path and name of each video (MOV, MP4, 3GP). | */home/user/videos/D29 IphoneX/IMG 2414.MOV* |
| *Path-origin* | Contains only the video path (MOV, MP4, 3GP). | */home/user/videos/D29 IphoneX* |
| *Class-label* | Indicates the ID of each class (brand, social network, instant messaging application, and editing program). | *1=Apple* |
| *File-name* | Represents the name of the video in MOV, MP4 and 3GP format. | *IMG 2414.MOV* |
| *Marker* | It is the brand of the device that generated the video. Metadata that usually insert Apple devices. | *Apple* |
| *Model* | It is the model of the device that generated the video. Metadata that usually insert Apple devices. | *Iphone X* |
| *PathOrder-tag* | Contains the atoms with its respective relative order of each level. Also, atoms with relative order followed of the tag. | */ftyp-1/* and */ftyp-1/majorBrand* |
| *Value* | These are the values of each *PathOrder-tag*. | *qt* |
| *Reading-orders* | It is the absolute order of each *PathOrder-tag*. Also called reading order. | *1* |

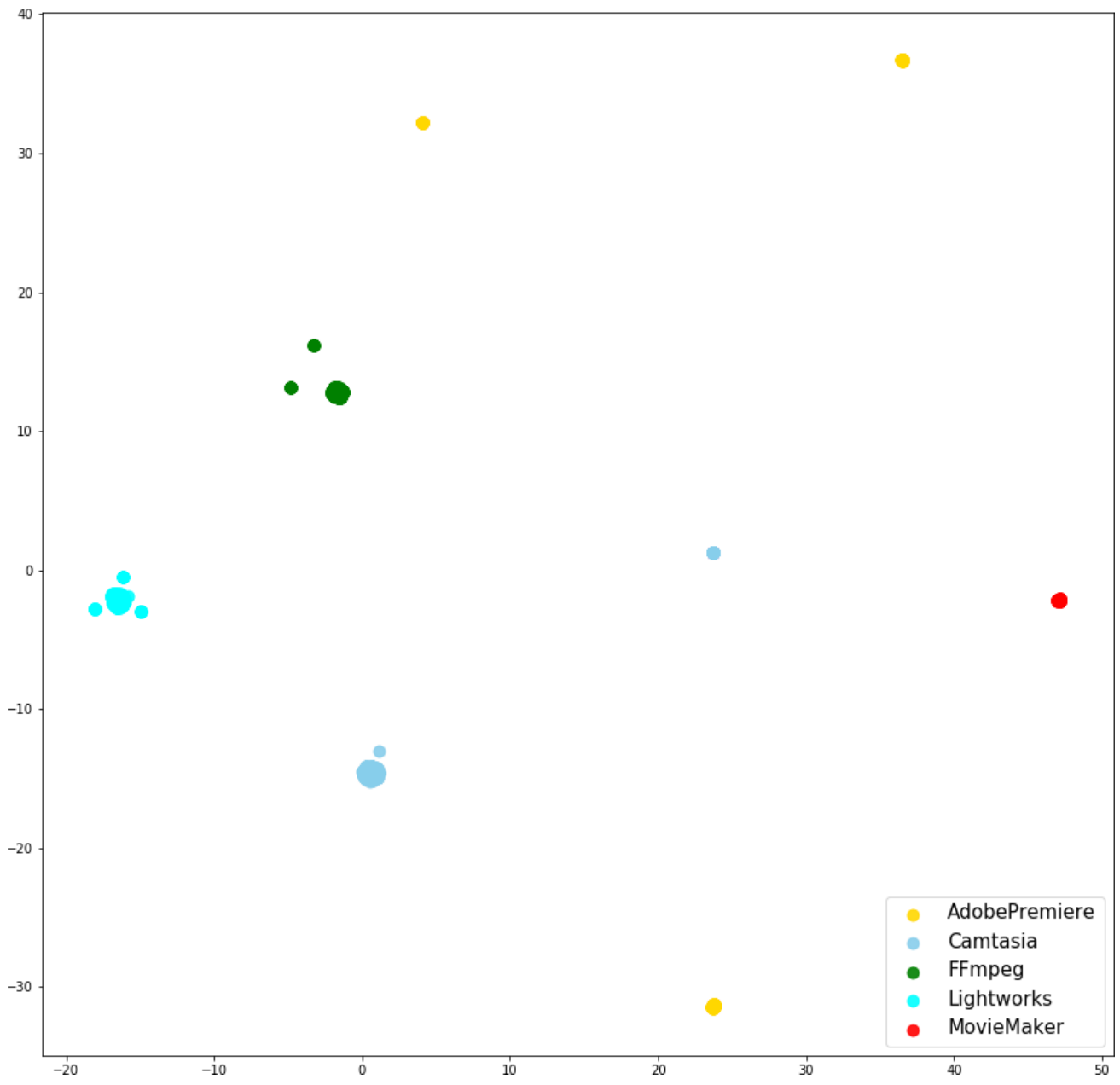

Figure 9: Visualization of the dataset (Editing programs) using the t-SNE

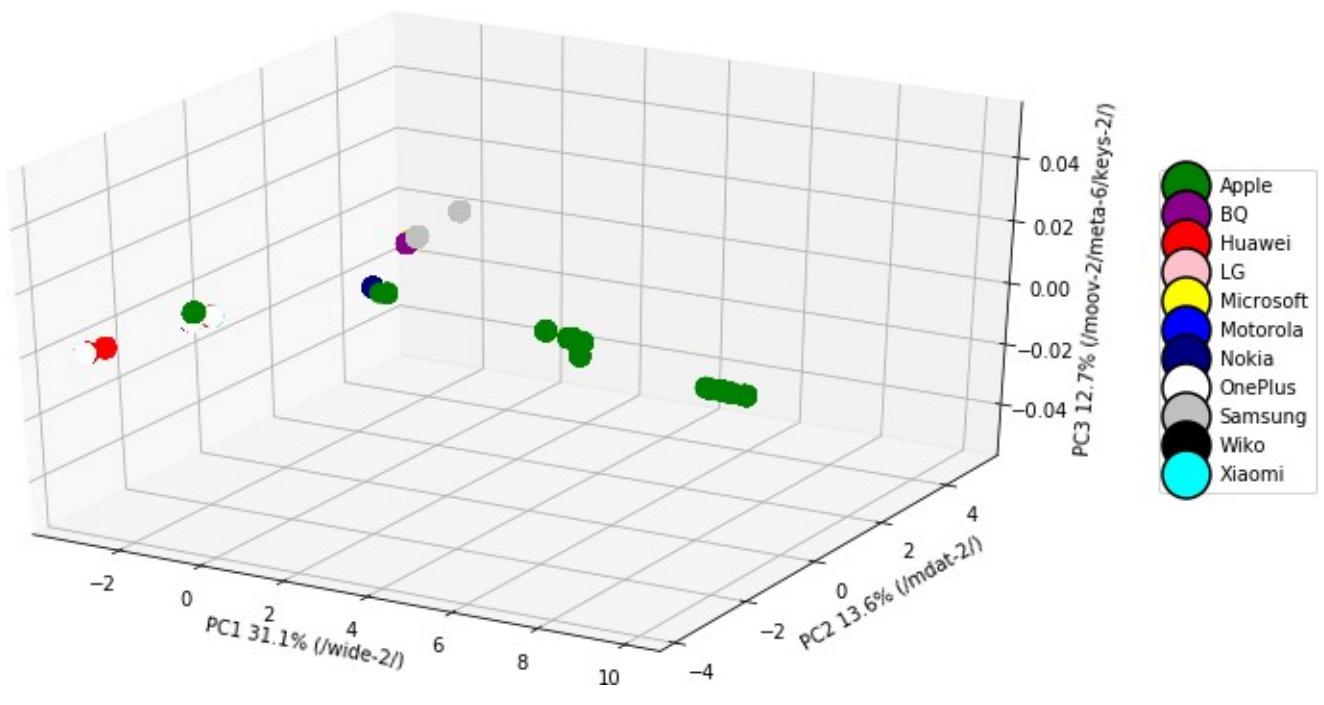

Figure 10: Principal component analysis - Original videos

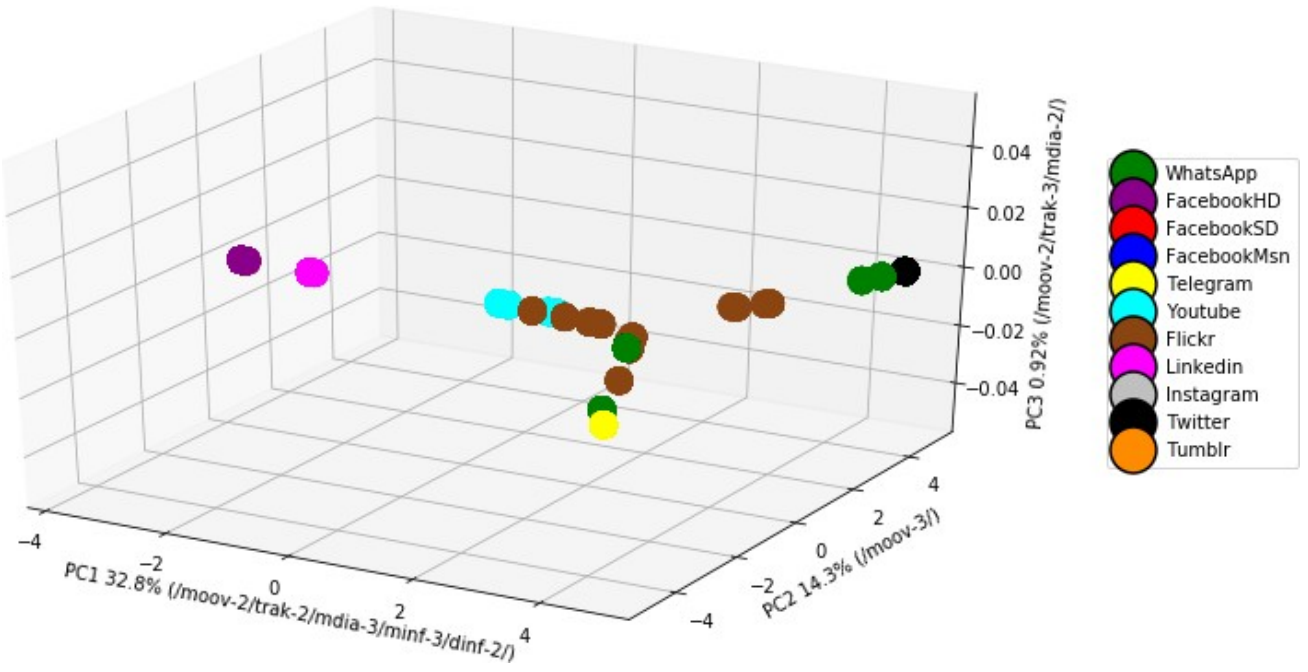

Figure 11: Principal component analysis - SN and IMA

*2/trak-2/mdia-3/minf-3/dinf-2/* contains 32.8% of the variance, the second */moov-3/* contains 14.3% and the third */moov-2/trak-3/mdia-2/* has 0.92%, adding up to 48.02%; while the remaining 21 components account for 50.98% of the variance. In Figure 11, the three-dimensional scatter plot is presented using the SN and IMA dataset. Likewise, the videos edited (721) by EP group a total of 332 dimensions, and by applying PCA with a configuration of 0.99 and 0.95, it was possible to reduce them to 7 and 6 dimensions, respectively. The first principal component */moov-2/trak-3/mdia-2/minf-3/stbl-3/stco-5/* contains 28.0% of the variance, the second */moov-2/trak-3* contains 25.1% and the third (/free-3/) has 17.0%, adding up to 70.1%; while the remaining 6 components account for 29.4% of the variance. In Figure 12, the three-dimensional scatter plot using the EP dataset is presented.

To reinforce the exploratory analysis, all unique *PathOrder-tag* were consolidated for each class and then compared to each other. The Pearson correlation coefficient was used, which represents the numerical expression that indicates the degree of correspondence or relationship that exists between two variables. These numbers vary between [-1, 1]. If the value obtained is "1", it indicates that the two classes are perfectly related. If the value is "0", it indicates that there is no linear relationship between them. If the value is negative, it indicates that there is a negative correlation. Therefore, the closer the correlation value gets to -1, the higher the difference between classes will be, and this is a positive precedent to take into account in the implementation of our proposal.

Figure 13 presents the correlation matrix between brands, where a perfect correlation is observed between videos of the same brands. Apple, Nokia, Microsoft, and BQ differ significantly from other brands. The brands (Huawei, Motorola), (Huawei, OnePlus), (Huawei, Xiaomi), (Motorola, Xiaomi), (OnePlus, Xiaomi) achieve a very high positive correlation.

Figure 14 presents the correlation matrix between social networks and instant messaging applications. We can see that there is only a perfect correlation between similar classes. However, there is a very high positive correlation between the classes (Facebook SD, Tumblr), (Facebook Msn, Instagram), (Facebook Msn, Tumblr), (Instagram, Tumblr). On the other hand, the most separate classes are WhatsApp, Telegram, Youtube, Twitter, and Flickr (the latter for not making any additional compression on the video).

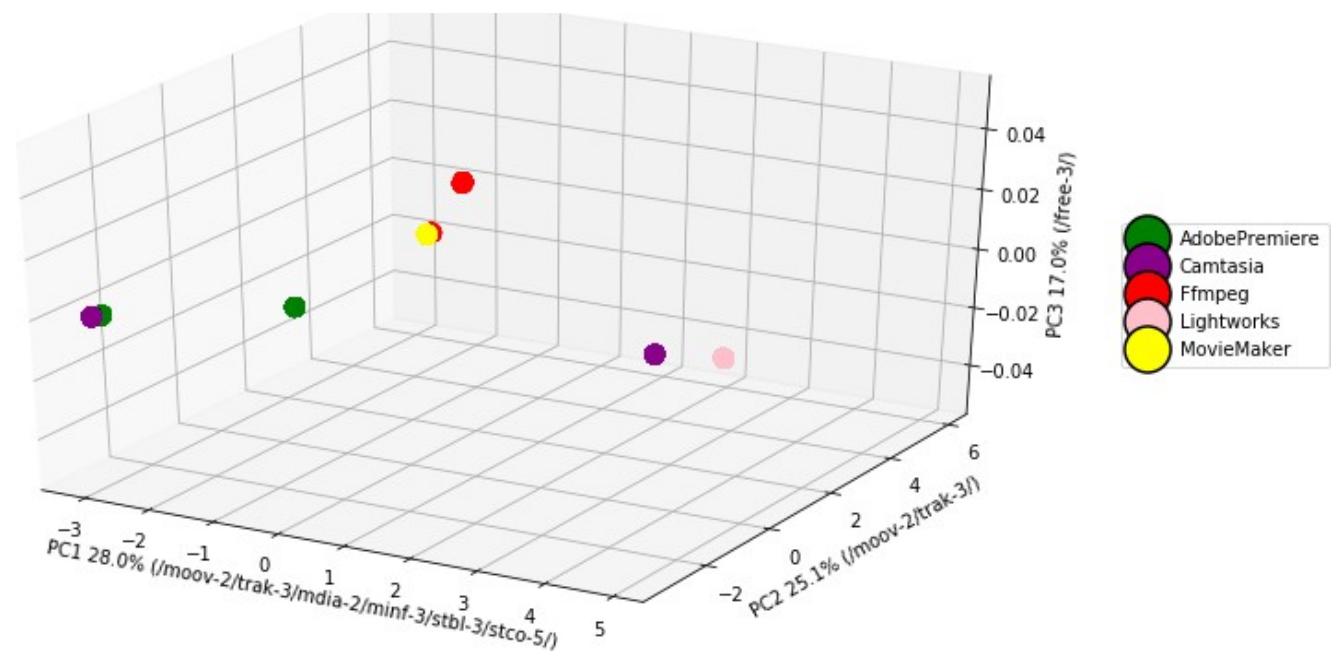

Figure 12: Principal component analysis - Editing programs

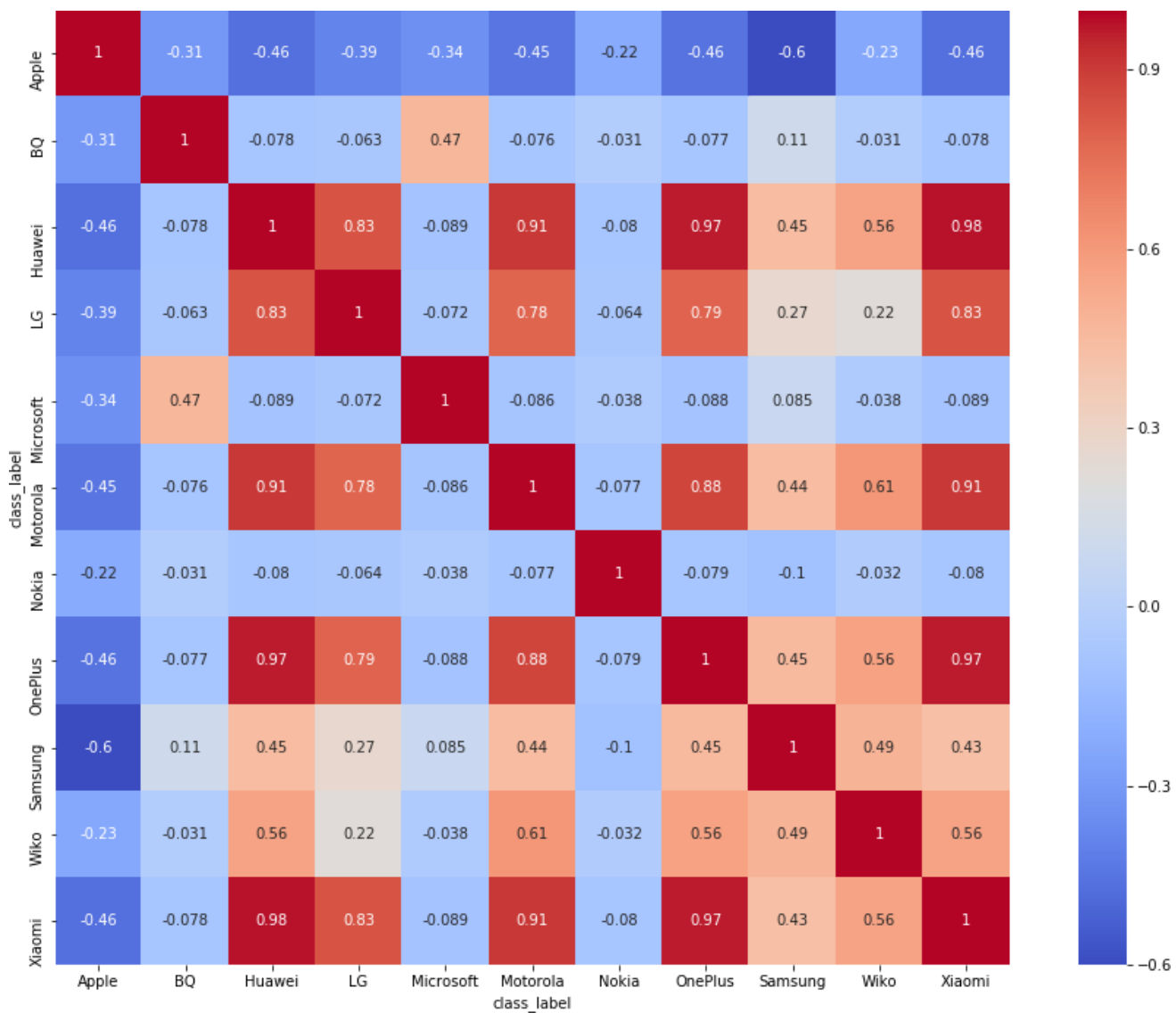

Figure 13: Correlation matrix between brands

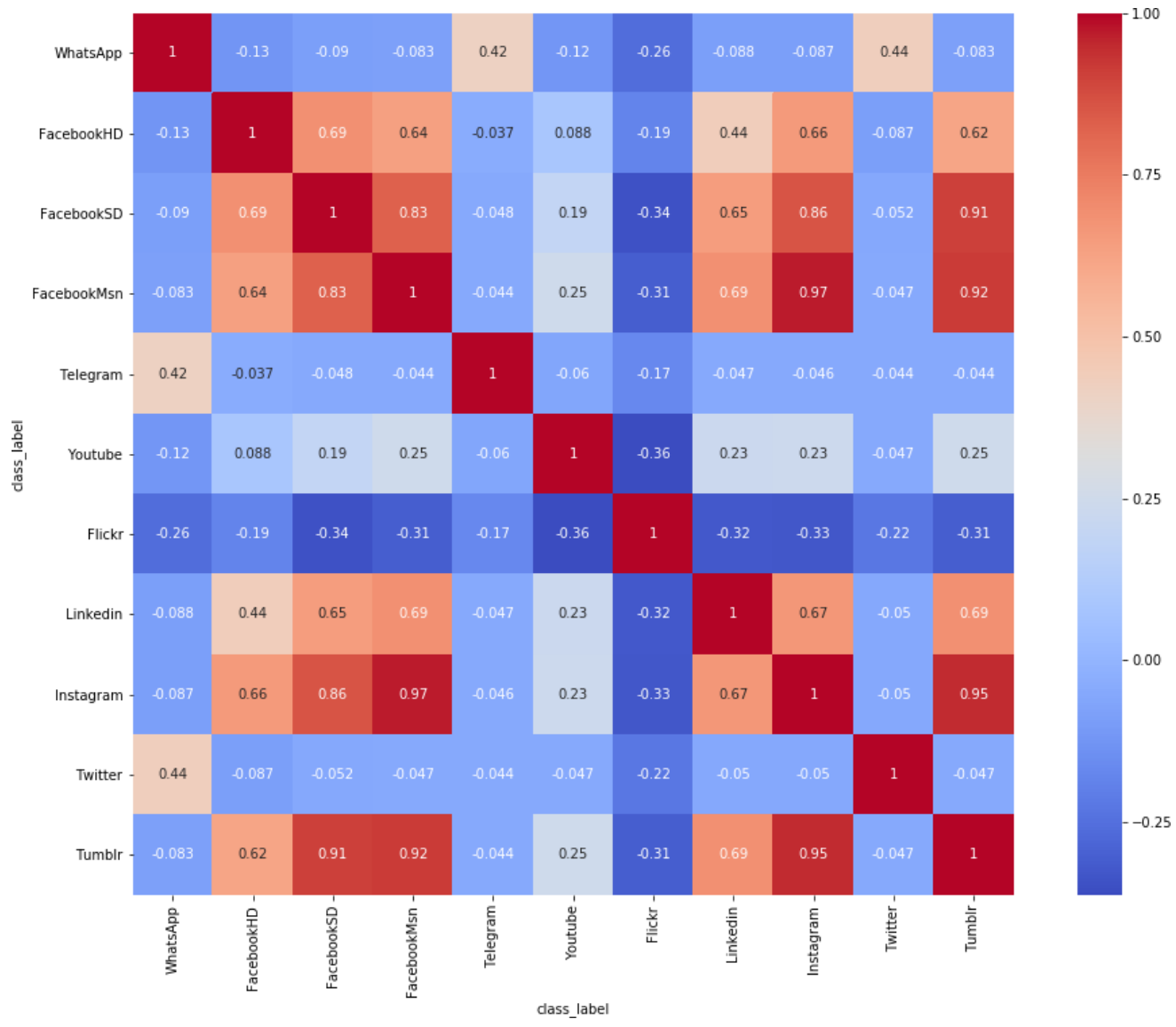

Figure 14: Correlation matrix between SN and IMA

Figure 15 presents the correlation matrix between the editing programs, and also shows that there is only a perfect correlation between similar classes. Likewise, all classes differ markedly from each other. To have a broader view of the data, Figure 16 presents the correlation matrix between the 24 classes that make up the dataset. Again, there is only a perfect correlation between similar classes. Also, the group of brands has a higher correlation concerning the group of social networks and editing programs. Finally, it is necessary to highlight the very high correlation between the classes (Nokia, Movie Maker), (Instagram, Tumblr).

After analyzing the data using t-SNE, Pearson correlation coefficient and PCA, we have been able to confirm that there are substantial differences between classes.

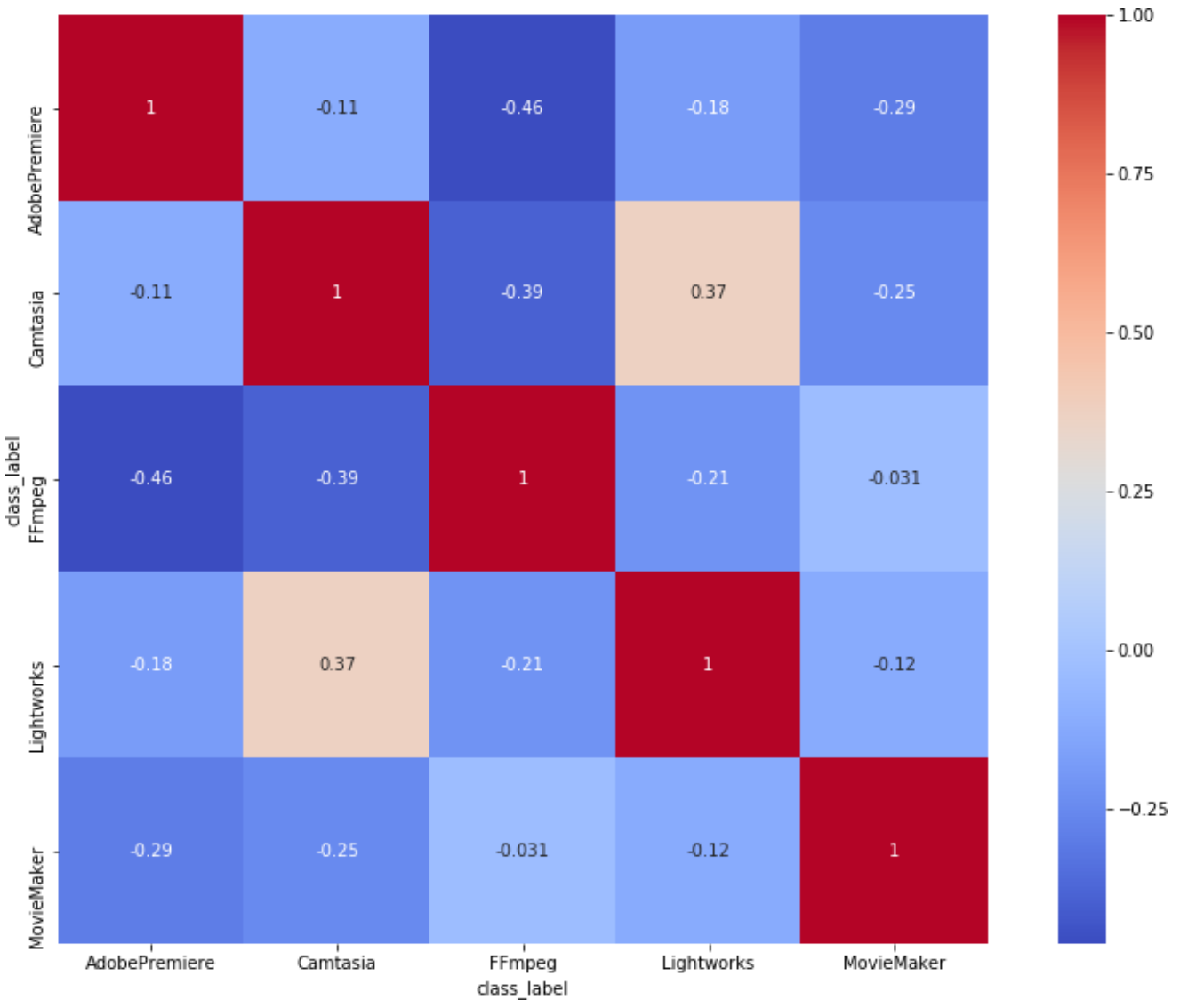

Figure 15: Correlation Matrix between editing programs

### 3.3.2. *Values and metadata analysis*

The values of the tags are classified into editable and non-editable values. Editable values, also called metadata, are those that are generally of the string type and are found in *meta, mebx, free, udta* atoms. These atoms can appear at all levels of the hierarchical structure of the multimedia container. The modification of these values does not influence the reproduction of videos. The *meta* atom contains child atoms that provide values with substantial information for analysis.

For example, an alternative for finding the brand name: *com. apple.quicktime.make: Apple* is to follow the path *moov* → *meta* → *keys* → *mdta*. To know if the video was captured with the front or back camera: *com.apple.quicktime. camera.identifier: Back* is to follow the path *moov* → *trak* →

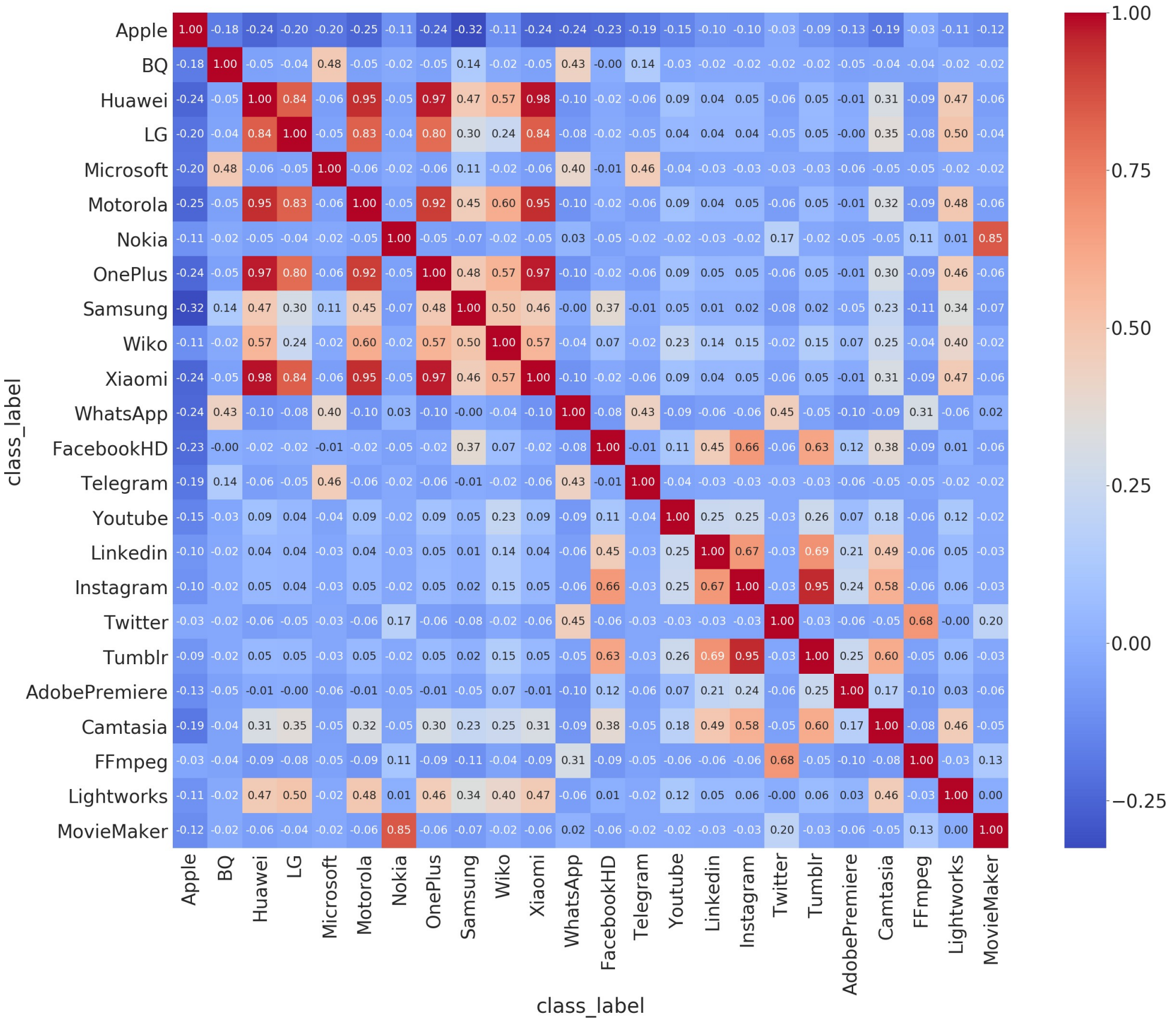

Figure 16: Correlation matrix between all classes

*meta* → *keys* → *mdta*. The *mebx* atom usually appears in the *trak-3* and *trak-4* atoms of videos generated by new Apple-branded mobile phones and is intended to store metadata information. Among the fundamental values that the *mebx* atom possesses is *com.apple.quicktime.video-orientation:90* and follow the next path: *moov* → *trak* → *mdia* → *minf* → *stbl* → *stsd* → *mebx* → *keys* → *local_key_id*1 → *keyd* → *keyValueArray*.

On the other hand, non-editable values are made up of bytes, and to display them. It is necessary to convert them into a hexadecimal value or any other type indicated by the specification. If an attacker deletes or modifies any of these values, it will undoubtedly affect the reproduction of the digital video. This means they are robust variables for the exploitation of forensic science. Among these values are *numOfSequenceParameterSetsPaddingBits: 7*, which is located in the path *moov* → *trak* → *mdia* → *minf* → *stbl* → *stsd* → *avc*1 → *avcC* and *objectTypeIndication: 64* of the path: *moov* → *trak* → *mdia* → *minf* → *stbl* → *stsd* → *mp4a* → *esds*.

Tables 9, 10, 11 and 12 present the analysis of the most representative editable and non-editable values of original and manipulated videos described in our datasets. Table 9 shows that Apple brand videos (Iphone XR) have more metadata information such as brand, model, and operating system version. The three values of the *timeScale* tags (600, 600, 44100) are very different from the other models. It also has the video compression codec (HEVC) and is the only model that inserts the Colour Parameter type with value *nclc*. On the other hand, Huawei P9, LG Nexus 5, Motorola Nexus 6, One Plus A3003, Samsung S9 Plus, and Xiaomi Redmi Note3 have the operating system version on the *com.android.version* tag. Samsung Galaxy S9 Plus is the only one that has values in the *com.android.capture.fps* tag, while Nokia 808 Pureview is the only one that does not have values in the *componentName* tag in the video track. Among the non-editable values are *componentSubtype:vide* (video track) and *componentSubtype:soun* (audio track).

Table 10 shows the values inserted by each social network and instant messaging application. Facebook HD and Facebook SD insert values similar to the tags, except for *timeScale*, *metadataCompressor*, *mediaWidth*, and *mediaHeight*. On the other hand, some applications insert different values according to the shared video format (MOV or MP4) such as Youtube, Linkedin, and WhatsApp. For example, Youtube embeds the value 15360 to the *timeScale* tag if the transferred video is MP4, and if it is MOV 1000. Linkedin, embed the value 1 in the *hSpacing* tag if it is MP4 and 405 if it is MOV. WhatsApp inserts the value of 1 to the *minorVersion* tag if the original video is MOV format and the value of 0 if it is MP4 or 3GP. Telegram and WhatsApp are the only ones that insert the value *nclx* in the *colorParameterType* tag. Tumblr adds the value 640 in the *hSpacing* and *vSpacing* tags. Finally, Youtube embeds the value of *Google* in the *metadataCompressor* tag, while WhatsApp and Telegram do not enter any value.

Table 9: Values and metadata - Brands

| | Brand | Apple | BQ | Huawei | LG | Microsoft | Motorola | Nokia | One Plus | Samsung | Wiko | Xiaomi |
|---|---|---|---|---|---|---|---|---|---|---|---|---|
| | **Tag/Model** | **Iphone XR** | **Aquaris4.5** | **P9** | **Nexus 5** | **Lumia640Lte** | **Nexus 6** | **808PureView** | **A3003** | **GalaxyS9Plus** | **Ridge4G** | **RedmiNote3** |
| **Format** | majorBrand | qt | mp42 | mp42 | mp42 | mp42 | mp42 | mp42 | mp42 | mp42 | mp42 | mp42 |
| | minorVersion | 0 | 0 | 0 | 0 | 0 | 0 | 0 | 0 | 0 | 0 | 0 |
| | compatibleBrands | qt | isom, mp42 | isom, mp42 | isom, mp42 | mp42, 3gp6, isom, M4V | isom, mp42 | mp42, 3gp4, isom, avc1 | isom, mp42 | isom, mp42 | isom, mp42 | isom, mp42 |
| **Global Tags** | timeScale | 600 | 1000 | 1000 | 1000 | 1000 | 1000 | 10000 | 1000 | 1000 | 1000 | 1000 |
| | nextTrackId | 5 | 3 | 3 | 3 | 3 | 3 | 3 | 3 | 3 | 3 | 3 |
| | preferredRate | 1 | 1 | 1 | 1 | 1 | 1 | 1 | 1 | 1 | 1 | 1 |
| | preferredVolume | 1 | 1 | 1 | 1 | 1 | 1 | 1 | 1 | 1 | 1 | 1 |
| | com.apple.quicktime.make | Apple | - | - | - | - | - | - | - | - | - | - |
| | com.apple.quicktime.model | iPhone XR | - | - | - | - | - | - | - | - | - | - |
| | com.apple.quicktime.software | 12.1.4 | - | - | - | - | - | - | - | - | - | - |
| | com.android.version | - | - | 6.0 | 6.0.1 | - | 6.0.1 | - | 7.0 | 8.0.0 | - | 6.0.1 |
| | com.android.capture.fps | - | - | - | - | - | - | - | - | 1114636288 | - | - |
| **Video Track** | timeScale | 600 | 90000 | 90000 | 90000 | 30000 | 90000 | 30000 | 90000 | 90000 | 90000 | 90000 |
| | graphicsMode | 64 | 0 | 0 | 0 | 0 | 0 | 0 | 0 | 0 | 0 | 0 |
| | componentSubtype | vide | vide | vide | vide | vide | vide | vide | vide | vide | vide | vide |
| | componentName | Core Media Data Handler | VideoHandle | VideoHandle | VideoHandle | VideoHandler | VideoHandle | - | VideoHandle | VideoHandle | VideoHandle | VideoHandle |
| | componentType | dhlr | - | - | - | - | - | - | - | - | - | - |
| | compressorName | HEVC | - | - | - | - | - | - | - | - | - | - |
| | extensions | hvcC | avcC, pasp | avcC, pasp | avcC, pasp | avcC | avcC, pasp | avcC | avcC, pasp | avcC, pasp | avcC, pasp | avcC, pasp |
| | hSpacing | - | 65536 | 65536 | 65536 | 65536 | 65536 | 65536 | 65536 | 65536 | 65536 | 65536 |
| | vSpacing | - | 65536 | 65536 | 65536 | 65536 | 65536 | 65536 | 65536 | 65536 | 65536 | 65536 |
| | spatialQuality | 512 | 0 | 0 | 0 | 0 | 0 | 0 | 0 | 0 | 0 | 0 |
| | verticalResolution | 72 | 72 | 72 | 72 | 72 | 72 | 72 | 72 | 72 | 72 | 72 |
| | horizontalResolution | 72 | 72 | 72 | 72 | 72 | 72 | 72 | 72 | 72 | 72 | 72 |
| | temporalQuality | 512 | 0 | 0 | 0 | 0 | 0 | 0 | 0 | 0 | 0 | 0 |
| | colorTableID | 65535 | 65535 | 65535 | 65535 | 65535 | 65535 | 65535 | 65535 | 65535 | 65535 | 65535 |
| | depth | 24 | 24 | 24 | 24 | 24 | 24 | 24 | 24 | 24 | 24 | 24 |
| | mediaWidth | 1920 | 1280 | 1920 | 1920 | 1920 | 1920 | 1920 | 1920 | 3840 | 1920 | 1920 |
| | mediaHeight | 1080 | 720 | 1080 | 1080 | 1080 | 1080 | 1080 | 1080 | 2160 | 1080 | 1080 |
| | colorParameterType | nclc | - | - | - | - | - | - | nclx | nclx | - | - |

| | Brand | Apple | BQ | Huawei | LG | Microsoft | Motorola | Nokia | One Plus | Samsung | Wiko | Xiaomi |
|---|---|---|---|---|---|---|---|---|---|---|---|---|
| | **Tag/Model** | **Iphone XR** | **Aquaris4.5** | **P9** | **Nexus 5** | **Lumia640Lte** | **Nexus 6** | **808PureView** | **A3003** | **GalaxyS9Plus** | **Ridge4G** | **RedmiNote3** |
| **Audio Track** | timeScale | 44100 | 48000 | 48000 | 48000 | 48000 | 48000 | 48000 | 48000 | 48000 | 48000 | 48000 |
| | componentSubtype | soun | soun | soun | soun | soun | soun | soun | soun | soun | soun | soun |
| | componentName | Core Media Audio | SoundHandle | SoundHandle | SoundHandle | AudioHandler | SoundHandle | SoundHandle | SoundHandle | SoundHandle | SoundHandle | SoundHandle |
| | componentType | mhlr | - | - | - | - | - | - | - | - | - | - |
| | sampleRate | 44100 | 48000 | 48000 | 48000 | 48000 | 48000 | 48000 | 48000 | 48000 | 48000 | 48000 |
| | compressionID | 65534 | 0 | 0 | 0 | 0 | 0 | 0 | 0 | 0 | 0 | 0 |
| | dataFormat | mp4a | mp4a | mp4a | mp4a | mp4a | mp4a | mp4a | mp4a | mp4a | mp4a | mp4a |
| | extensions | esds | esds | esds | esds | esds | esds | esds | esds | esds | esds | esds |
| | bytesPerFrame | 2 | - | - | - | - | - | - | - | - | - | - |
| | numOfChannels | 2 | 2 | 2 | 1 | 1 | 1 | 2 | 2 | 2 | 2 | 2 |
| | bytesPerPacket | 1 | 1 | 1 | 1 | 1 | 1 | 1 | 1 | 1 | 1 | 1 |
| | bytesPersample | 2 | 2 | 2 | 2 | 2 | 2 | 2 | 2 | 2 | 2 | 2 |
| | revisionLevel | 0 | 0 | 0 | 0 | 0 | 0 | 0 | 0 | 0 | 0 | 0 |
| | dataReferenceIndex | 1 | 1 | 1 | 1 | 1 | 1 | 1 | 1 | 1 | 1 | 1 |
| **Metadata Tracks** | componentSubtype | meta | - | - | - | - | - | - | - | - | - | - |
| | componentName | Core Media Metadata | - | - | - | - | - | - | - | - | - | - |
| | componentManufacturer | appl | - | - | - | - | - | - | - | - | - | - |
| | mebx/roll-angle | 0 | - | - | - | - | - | - | - | - | - | - |
| | mebx/face.face-id | 0 | - | - | - | - | - | - | - | - | - | - |
| | mebx/face.bounds | 0 | - | - | - | - | - | - | - | - | - | - |
| | mebx/video-orientation | 90 | - | - | - | - | - | - | - | - | - | - |
| | balance | 0 | - | - | - | - | - | - | - | - | - | - |
| | stsd tmcd | False | - | - | - | - | - | - | - | - | - | - |
| | componentFlags | 0 | - | - | - | - | - | - | - | - | - | - |

Table 10: Values and metadata - Social networks and Instant messaging applications

| | SN/IMA Tag | Facebook HD | Facebook SD | Youtube | Linkedin | Instagram | Twitter | Tumblr | Facebook Msn | WhatsApp | Telegram |
|---|---|---|---|---|---|---|---|---|---|---|---|
| **Format** | majorBrand | isom | isom | mp42 | isom | isom | isom | isom | isom | mp42 | mp42 |
| | minorVersion | 512 | 512 | 0 | 512 | 512 | 512 | 512 | 512 | 0, 1 | 1 |
| | compatibleBrands | isom, iso2 avc1, mp41 | isom, iso2 avc1, mp41 | isom, mp42 | isom, iso2 avc1, mp41 | isom, iso2 avc1, mp41 | isom, iso2 avc1, mp41 | isom, iso2 avc1, mp41 | isom, iso2 avc1, mp41 | mp41, mp42, isom | mp41, mp42, isom |
| **Global Tags** | timeScale | 1000 | 1000 | 15360, 1000 | 1000 | 1000 | 1000 | 1000 | 1000 | 48000, 44100 | 44100 |
| | nextTrackId | 3 | 3 | 3 | 3 | 3 | 3 | 3 | 3 | 3 | 3 |
| | preferredRate | 1 | 1 | 1 | 1 | 1 | 1 | 1 | 1 | 1 | 1 |
| | preferredVolume | 1 | 1 | 1 | 1 | 1 | 1 | 1 | 1 | 1 | 1 |
| | reserved | 0 | 0 | 0 | 0 | 0 | 0 | 0 | 0 | 0 | 0 |
| | posterTime | 0 | 0 | 0 | 0 | 0 | 0 | 0 | 0 | 0 | 0 |
| **Video Track** | timeScale | 15360 | 90000 | 15360 | 15360 | 15360 | 90000 | 15360 | 15360 | 30057, 600 | 600 |
| | graphicsMode | 0 | 0 | 0 | 0 | 0 | 0 | 0 | 0 | 0 | 0 |
| | componentSubtype | vide | vide | vide | vide | vide | vide | vide | vide | vide | vide |
| | componentName | VideoHandler | VideoHandler | ISO Media file produced by Google Inc | VideoHandler | VideoHandler | VideoHandler | VideoHandler | VideoHandler | Core Media Video | Core Media Video |
| | componentType | - | - | - | - | - | - | - | - | - | - |
| | metadataCompressor | Lavf57.71.100 | Lavf56.40.101 | Google | Lavf57.71.100 | Lavf56.40.101 | Lavf57.66.105 | Lavf56.25.101 | Lavf56.40.101 | - | - |
| | extensions | avcC | avcC | avcC | avcC | avcC | avcC | avcC | avcC | avcC | avcC |
| | hSpacing | - | - | - | 1, 405 | - | - | 640 | - | - | 3 |
| | vSpacing | - | - | - | 1, 404 | - | - | 640 | - | - | 3 |
| | spatialQuality | 0 | 0 | 0 | 0 | 0 | 0 | 0 | 0 | 0 | 0 |
| | verticalResolution | 72 | 72 | 72 | 72 | 72 | 72 | 72 | 72 | 72 | 72 |
| | horizontalResolution | 72 | 72 | 72 | 72 | 72 | 72 | 72 | 72 | 72 | 72 |
| | temporalQuality | 0 | 0 | 0 | 0 | 0 | 0 | 0 | 0 | 0 | 0 |
| | colorTableID | 65535 | 65535 | 65535 | 65535 | 65535 | 65535 | 65535 | 65535 | 65535 | 65535 |
| | depth | 24 | 24 | 24 | 24 | 24 | 24 | 24 | 24 | 24 | 24 |
| | mediaWidth | 1280 | 400 | 1280 | 1280 | 480 | 720 | 852 | 1920 | 848 | 848 |
| | mediaHeight | 720 | 224 | 720 | 720 | 480 | 1280 | 480 | 1080 | 480 | 464 |
| | colorParameterType | - | - | - | - | - | - | - | - | nclx | nclx |
| **Audio Track** | timeScale | 44100 | 44100 | 44100 | 48000 | 48000 | 48000 | 48000 | 48000 | 48000 | 48000 |
| | componentSubtype | soun | soun | soun | soun | soun | soun | soun | soun | soun | soun |
| | componentName | SoundHandler | SoundHandler | ISO Media file produced by Google Inc. | SoundHandler | SoundHandler | SoundHandler | SoundHandler | SoundHandler | Core Media Audio | Core Media Audio |
| | componentType | - | - | - | - | - | - | - | - | - | - |
| | sampleRate | 48000, 44100 | 48000, 44100 | 44100 | 48000 | 48000, 44100 | 48000, 44100 | 48000 | 48000, 44100 | 48000, 44100 | 44100 |
| | compressionID | 0 | 0 | 0 | 0 | 0 | 0 | 0 | 0 | 0 | 0 |
| | dataFormat | mp4a | mp4a | mp4a | mp4a | mp4a | mp4a | mp4a | mp4a | mp4a | mp4a |
| | extensions | esds | esds | esds | esds | esds | esds | esds | esds | esds | esds |
| | bytesPerFrame | - | - | - | - | - | - | - | - | - | - |
| | numOfChannels | 2 | 2 | 1, 2 | 2 | 2 | 2 | 2 | 2 | 2 | 2 |
| | bytesPerPacket | - | - | - | - | - | - | - | - | - | - |
| | bytesPersample | - | - | - | - | - | - | - | - | - | - |
| | revisionLevel | 0 | 0 | 0 | 0 | 0 | 0 | 0 | 0 | 0 | 0 |
| | dataReferenceIndex | 1 | 1 | 1 | 1 | 1 | 1 | 1 | 1 | 1 | 1 |

Table 11: Values and metadata - Editing programs

| Tag/Editing program | | Adobe Premiere | Camtasia | FFmpeg | Lightworks | Movie Maker |
|---|---|---|---|---|---|---|
| **Format** | majorBrand | qt, mp42 | mp42 | qt, mp42 | mp42 | mp42 |
| | minorVersion | 537199360 | 0 | 512 | 0 | 0 |
| | compatibleBrands | qt, isom mp42 | isom, mp42 | qt, isom mp42 | isom, mp42 | mp41, isom |
| **Global Tags** | timeScale | 30000 | 90000 | 1000 | 90000 | 48000 |
| | nextTrackId | 4 | 3 | 3 | 3 | 3 |
| | preferredRate | 1 | 1 | 1 | 1 | 1 |
| | preferredVolume | 1 | 1 | 1 | 1 | 1 |
| | selectionTime | 0 | 0 | 0 | 0 | 0 |
| | previewDuration | 0 | 0 | 0 | 0 | 0 |
| | selectionDuration | 0 | 0 | 0 | 0 | 0 |
| | componentSubtype | - | mdir | - | - | mdir |
| **Video Track** | timeScale | 30000 | 30000 | 19200 | 30000 | 29970 |
| | graphicsMode | 64 | 0 | 0 | 0 | 0 |
| | componentHdlrName | Apple Alias Data Handler | - | DataHandler | - | - |
| | componentHdlr | alis | - | url | - | - |
| | componentName | Apple Video Media Handler | Mainconcept MP4 Video Media Handler | VideoHandler | Mainconcept MP4 Video Media Handler | VideoHandler |
| | componentType | mhlr, dhlr | - | mhlr, dhlr | - | - |
| | componentSubtype | vide | vide | vide | vide | vide |
| | compressorName | DV25 NTSC | VC Coding | HEVC, H.264 | VC Coding | VC Coding |
| | extensions | pasp, fiel | avcC | hvcC, avcC | avcC | avcC |
| | hSpacing | 872 | - | - | - | - |
| | vSpacing | 720 | - | - | - | - |
| | spatialQuality | 1023 | 0 | 512 | 0 | 0 |
| | verticalResolution | 72 | 72 | 72 | 72 | 72 |
| | horizontalResolution | 72 | 72 | 72 | 72 | 72 |
| | temporalQuality | 0 | 0 | 512 | 0 | 0 |
| | colorTableID | 65535 | 65535 | 65535 | 65535 | 65535 |
| | depth | 24 | 24 | 24 | 24 | - |
| | mediaWidth | 720 | 1280 | 1920 | 1280 | 1920 |
| | mediaHeight | 480 | 720 | 1080 | 720 | 1080 |
| | colorParameterType | nclc | - | - | - | - |

Table 12: Values and metadata - Editing programs

| | Tag/Editing program | Adobe Premiere | Camtasia | FFmpeg | Lightworks | Movie Maker |
|---|---|---|---|---|---|---|
| **Audio Track** | timeScale | 48000 | 44100 | 44100 | 48000 | 48000 |
| | componentHdlrName | Apple Alias Data Handler | - | DataHandler | - | - |
| | componentHdlr | alis | - | url | - | - |
| | componentName | Apple Sound Media Handler | Mainconcept MP4 Sound Media Handler | SoundHandler | Mainconcept MP4 Sound Media Handler | SoundHandler |
| | componentType | mhlr, dhlr | - | mhlr, dhlr | - | - |
| | componentSubtype | soun | soun | soun | soun | soun |
| | sampleRate | 48000 | 44100 | 44100 | 48000 | 48000 |
| | compressionID | 65535 | 0 | 65534 | 0 | 0 |
| | dataFormat | sowt | mp4a | mp4a | mp4a | mp4a |
| | extensions | chan | esds | esds | esds | esds |
| | bytesPerFrame | 4 | - | 0 | - | - |
| | numOfChannels | 2 | 2 | 1 | 2 | 2 |
| | bytesPerPacket | 2 | - | 0 | - | - |
| | bytesPersample | 2 | - | 2 | - | - |
| | revisionLevel | 0 | 0 | 0 | 0 | 0 |
| | dataReferenceIndex | 1 | 1 | 1 | 1 | 1 |
| **Metadata Tracks** | componentSubtype | meta | - | - | - | - |
| | componentName | Core Media Metadata | - | - | - | - |
| | componentManufacturer | appl | - | - | - | - |
| | mebx/roll-angle | 0 | - | - | - | - |
| | mebx/face.face-id | 0 | - | - | - | - |
| | mebx/face.bounds | 0 | - | - | - | - |
| | mebx/video-orientation | 90 | - | - | - | - |
| | balance | 0 | - | - | - | - |
| | stsd tmcd | False | - | - | - | - |
| | componentFlags | 0 | - | - | - | - |

Tables 11 and 12 show that Camtasia, Lightworks, and Movie Maker insert the same pattern to MP4 and MOV videos, while Adobe Premier and FFmpeg add different values to *majorBrand, compatibleBrands, componentHdlrName, componentName* tags. Camtasia and Lightworks have similar value tags, such as *timeScale, componentName, compressorName*, and differ in the value of the *componentSubtype* tag. FFmpeg inserts very different values in *componentHdlr* and *componentType* tags and is the only editor that maintains the original video codec type (HEVC or H.264) and is due to the kind of manipulation that was performed (Copy -vcodec).

## 4. Video Container Counter-Forensics

The proposed analysis has also determined that the multimedia container structure can be unconsciously manipulated only by sharing the videos through a social network or messaging application. Also, people with little knowledge of video editing could use some software to improve the quality of the scene or add some effects.

However, malicious people with a stronger knowledge of video manipulation could make use of techniques called counter-forensics to hinder the process of forensic analysis [38] [39] [40]; precisely, they can manipulate atoms, remove metadata and insert information into the atoms by performing the following actions:

- The attacker could use some hex editor to edit the name of atoms (4-byte ASCII code) considered not "necessary" for the video reproduction like *free, thkd, wide, meta, data, and udta*. Also, he could remove or edit some metadata without causing any damage to the video, for example, the metadata of the *data* tags that indicate the make and model of the mobile device. However, there are a lot of values that are intrinsic to each tag and are not editable such as *configurationVersion, AVCProfileIndication, AVCLevelIndication, lenghtSizeMinusOne*.

- The attacker could re-order the appearance of the main atoms in the multimedia container; this is because the order of appearance of the root atoms is variable *mdat, moov, wide, udta, free*, except *ftyp* which according to the specification is the one that provides the format of the file and must necessarily be the header of the multimedia container. However, this task is not easy, since changing the position of the *mdat* atom, which stores audio and video samples also imply adjusting the chunk offset tables *stco or co64*, which are those that describe the location of the samples and whose parent atoms are: $moov \rightarrow trak \rightarrow mdia \rightarrow minf \rightarrow stbl$.

- The attacker could insert all kinds of encrypted or unencrypted information (images, text, and so on) into a large number of *flags* tags or other tags with similar characteristics. Currently, there are video steganography tools such as *OpenPuff* [41] that can securely hide information by replacing the value of the *flags* tags with other encrypted data. However, in [42], the authors propose a method to detect the presence of OpenPuff video steganography in MP4 files, achieving positive results. Our analysis proposal is also suitable for detecting anomalies in the values of *flags* tags and others that may be prone to attacks by this type of technique in videos with MP4, MOV, and 3GP format. The original Apple brand videos have between 61 and 65 *flags* tags and have as values in hexadecimal (0, 7, 15). Social networks and Instant messaging applications, reduce the number of *flags* to 25, and the values inserted in hexadecimal are (0, 3, 1). The editing programs insert between 23 and 32 *flags*, and their values are much more varied, and among them are (0, 1, 7, 11, 12, 26, 28).

  We intentionally hid a 217.1 KB text file in 10 original Apple-branded videos using the *OpenPuff* tool and then extracted 100% of the features. The analysis found very high values in hexadecimal and, above all, very variable values. These values are between 0 and 13697024. Compared to the universal values of the original and manipulated video *flags*, there is a big difference. This analysis shows that these videos have hidden and encrypted elements that should be evaluated with more robust steganography techniques.

In summary, the analysis of the editable and non-editable values of each tag is fundamental to neutralize possible attacks and be more accurate in verifying the integrity and authenticity of digital videos.

## 5. Conclusions

This paper presents a method that exhaustively analyzes the intrinsic structure of multimedia containers. The technique allows verifying the integrity and authenticity of videos in MP4, MOV and 3GP format. Concretely, it can determine the brand of the mobile device that generated the video, the social network or instant messaging application that was used to transfer it, and the editing program that was used to execute some manipulation. Additionally, the method allows for the identification of anomalies in the values of each tag, caused by steganography tools. To this end, we first studied the technical specifications that define and describe the criteria to be used by manufacturers and technology platforms, to perform the processing of digital videos. Secondly, a sufficiently large dataset was generated, consisting of 1957 original videos, captured by 86 mobile devices, 66 models and 11 brands. Besides, we shared a total of 2598 videos through 8 social networks and 1084 videos through 3 instant messaging applications. Also, we manipulated 721 videos with 5 of the most used editing programs at present. Thirdly, the multimedia container structures were extracted with the atom extraction algorithm, resulting in a set of features called *PathOrder-tag*. Finally, the multimedia container structure analysis was carried out, which has been distributed in 3 parts: preliminary analysis, massive analysis, value and metadata analysis, including possible attacks (counter-forensic).

In the preliminary analysis, we compared the structure of an original video with the structure of the same video after being shared by three social networks and two messaging applications. This analysis shows that an original video changes its original structure after being shared by some social platform (except Flickr). Likewise, it was observed that each platform inserts a unique pattern that allows easy recognition of its origin. For the massive analysis, we used the absence and presence of the variables *PathOrder-tag* and applied the techniques t-SNE, PCA, and the Pearson Correlation Coefficient. When applying t-SNE, it was observed that the classes (brands, social platforms and editing program) are considerably separated from each other, although Apple differs significantly from other brands. When applying PCA, with the configuration (component parameter = 99), it was observed that the brands only require 20 dimensions to explain the total variance, 24 for social platforms and 7 for editing programs. When applying the Pearson Correlation Coefficient, the mobile device brands achieved a medium-high correlation between them, except Nokia and Microsoft. Social networks reached a medium-level correlation, except Facebook HD and Facebook SD. The editing programs reached a very low correlation, indicating that they have more considerable differences between them.

In the analysis of values and metadata, it was determined that Apple-branded devices have a higher amount of metadata because current models have up to 4 *trak* atoms. The social networks Facebook SD and Facebook HD insert similar values to almost all tags, and this is because both belong to the same technology group. Likewise, the method was evaluated against attacks that use steganography techniques such as the OpenPuff tool, managing to describe anomalies introduced into the multimedia container efficiently.

In summary, the structures and values of multimedia containers are patterns that allow determining the integrity and authenticity of digital videos (MP4, MOV, 3GP).

## Acknowledgements

This project has received funding from the European Unions Horizon 2020 research and innovation programme under grant agreement No 700326. Website: http://ramses2020.eu